\documentclass[12pt,preprint]{aastex}
\usepackage{amsmath,longtable,natbib}
\usepackage{lineno}
\usepackage{float}
\usepackage[caption = false]{subfig}
\usepackage{hyperref}
\usepackage{graphicx}
\shorttitle{GHRSS $-$ description and discovery}
\shortauthors{Bhattacharyya et al.}
\usepackage{subfig}

\begin{document}

\title{The GMRT High Resolution Southern Sky Survey for pulsars and transients -I. Survey description and initial discoveries}
\author{
B.~Bhattacharyya\altaffilmark{1},
S.~Cooper\altaffilmark{1},
M.~Malenta\altaffilmark{1},
J.~Roy\altaffilmark{1,2},
J.~Chengalur\altaffilmark{2},
M.~Keith\altaffilmark{1},
S.~Kudale\altaffilmark{2},
M.~McLaughlin\altaffilmark{3},
S.~M.~Ransom\altaffilmark{4},
P.~S.~Ray\altaffilmark{5},
B.~W.~Stappers\altaffilmark{1}
}
\altaffiltext{1}{Jodrell Bank Centre for Astrophysics, School of Physics and Astronomy, The University of Manchester, Manchester M13 9PL, UK}
\altaffiltext{2}{National Centre for Radio Astrophysics, Tata Institute of Fundamental Research, Pune 411 007, India}
\altaffiltext{3}{Department of Physics \& Astronomy, West Virginia University, Morgantown, WV 26506, US}
\altaffiltext{4}{National Radio Astronomy Observatory(NRAO), Charlottesville, VA 22903, USA}
\altaffiltext{5}{Space Science Division, Naval Research Laboratory, Washington, DC 20375-5352, USA}

\affil{}

\begin{abstract}

We are conducting a survey for pulsars and transients using the Giant Metrewave Radio Telescope (GMRT).
The GMRT High Resolution Southern Sky (GHRSS) survey is an off-Galactic-plane ($|b|$~$>$~5) survey in the declination 
range $-$40\degr~to $-$54\degr~at 322 MHz. With the high time (up to 30.72 $\mu$s) and frequency (up to 0.016275 MHz) 
resolution observing modes, the 5$\sigma$ detection limit is 0.5 mJy for a 2 ms pulsar with 10\% duty cycle at 322 MHz. 
Total GHRSS sky coverage of 2866 deg$^2$, will result from 1953 pointings, each covering 1.8 deg$^2$. The 10$\sigma$ 
detection limit for a 5 ms transient burst is 1.6 Jy for the GHRSS survey. In addition, the GHRSS survey can reveal 
transient events like rotating radio transients or fast radio bursts. With 35\% of the survey completed 
(i.e. 1000 deg$^2$), we report the discovery of 10 pulsars, one of which is a millisecond pulsar (MSP), one 
of the highest pulsar per square degree discovery rates for any off-Galactic plane survey. We re-detected 23 known in-beam 
pulsars. Utilising the imaging capability of the GMRT we also localised 4 of the GHRSS pulsars (including the MSP) 
in the gated image plane within $\pm$ 10\arcsec. We demonstrated rapid convergence in pulsar timing with a more precise 
position than is possible with single dish discoveries. We also exhibited that we can localise the brightest transient 
sources with simultaneously obtained lower time resolution imaging data, demonstrating a technique that may have application in the SKA.

\end{abstract}

\vskip 0.6 cm

\section{Introduction}
\label{sec:intro}
Pulsars are rapidly rotating neutron stars emitting a beam of radio waves from their magnetic poles, that sweep our
line-of-sight at the spin period.
In the 48 years since the discovery of pulsars, the present population of about 2300 known pulsars have spin 
periods ranging from 1.4 ms to 8.5 s, magnetic field strengths from 6.6$\times 10^{7}$ G to 1.2$\times 10^{13}$ G, 
 line-of-sight electron column density values, called dispersion measures (DMs), from 2.38 pc~cm$^{-3}$ to 
 1456 pc~cm$^{-3}$, flux densities at 400 MHz ranging from 0.1 mJy to 5000 mJy and
characteristic ages ranging from 2.2$\times 10^{2}$ years to 6.7$\times10^{10}$ years\footnote{http://www.atnf.csiro.au/people/pulsar/psrcat/}. 
In an attempt to understand the distribution of sources within the limits of the known population and to probe the 
limits themselves there are many ongoing surveys that are discovering pulsars at an encouraging rate.   
Although many of the newly discovered pulsars have similar properties to the known population, some of the new 
ones are pushing the boundary of known parameter space. For example Low-mass X-ray binary$-$radio millisecond pulsar 
transitioning system: J1023$+$0038 \citep{archibald09} and J1227$-$4853 \citep{roy15}, a hierarchical stellar triple pulsar 
system: J0337+1715 \citep{ransom14}, a massive neutron star$-$white dwarf binary: J0348$+$0432 \citep{antoniadis13}, and 
 a pulsar with a very dense planetary mass companion: J1719$-$1438 \citep{bailes11}. The diversity of properties of 
these objects justifies the requirement of having more sensitive surveys for further discoveries. 

Modeling by \cite{faucher06} indicates that the Galaxy contains about 10$^5$ pulsars and the number of presently known 
pulsars is only around 1\% of this population. This implies that a vast majority of pulsars are waiting to be discovered. 
Studies of pulsars yield a better understanding of 
a variety of physics problems, from acceleration of particles in ultra strong magnetic fields (primarily via study of emission 
properties of normal pulsars, having spin period $>$~30 ms) to probes of ultra dense matter (mostly via studying the timing properties of 
millisecond pulsars, having spin period $<$30 ms, that are very stable rotators). Pulsars provide useful probes of their environments, 
e.g. inside pulsar wind nebulae, the centre of Globular clusters, or the Galactic centre. Some of the normal pulsars are 
useful for investigation of single pulse behaviour and can exhibit interesting individual properties, like glitches \citep{lyne96}, 
profile state changes \citep{lyne10}, nulling \citep{backer70} and intermittency \citep{kramer06}.
The emission mechanism for pulsar radiation at radio wavelengths is not yet understood, and a range of theoretical models 
have been proposed (e.g. \cite{ruderman75, gil03, alexander14}) to explain it.
 The discovery of a new sample of pulsars  with diverse observational properties will provide additional constraints on the possible emission mechanism. Additional 
pulsars distributed in our Galaxy will aid the investigation of properties of the interstellar medium via scattering measurements 
as well as via dispersion measure and rotation measure studies.
The fractional rotational stability of millisecond pulsars (MSPs), one part in 10$^{15}$, is comparable to atomic clocks \citep{lorimer04}.
Such rotational stability, compactness second only to black holes, and their presence in binary systems, make MSPs ideal
laboratories to test the physics of gravity and as detectors for long-wavelength gravitational waves \citep{Detweiler79}.  
Some relativistic binary MSPs are useful for tests of gravity and an array of spatially distributed MSPs can be used to 
detect gravitational waves \citep{lee12}. In addition to the contribution towards the detection of gravitational wave signal, 
MSP evolutionary processes can be tracked through individual interesting discoveries of MSPs in special evolutionary phases. 
For example, the black-widow systems \citep{roberts11} provides a missing link between the binary and isolated MSPs and 
Low-mass X-ray binary$-$radio MSP transitioning redback systems \citep{archibald09,roy15} provides a probe of binary evolution. 

In addition to the regular emission from pulsars, a wide variety of transient phenomena are known at faster time scales 
(micro-seconds to seconds). The discoveries of giant pulses from pulsars \citep{lundgren95}, quasi-periodic emission from 
rotating radio transients (RRATs; \cite{McLaughlin06}), bursts of periodic pulsations from magnetars \citep{camilo06}
and highly dispersed fast radio bursts (FRBs) of possible extragalactic origin \citep{lorimer07} are a few examples.
The FRBs have been discovered in radio pulsar surveys at Parkes and Arecibo in the last decade \citep{lorimer07,thornton13,spitler14,burke14,petroff14}. 
All FRBs discovered to date have been single radio events of millisecond duration {with DM values generally higher 
than the possible Galactic contribution.} Majority of the FRB population theories suggest that they originate at cosmological 
distances (\cite{deng14,luan14,keane15a}), although the actual progenitors are still unknown.
 The probable extragalactic origin allows these bursts to be used for 
determination of the baryon content of the intergalactic medium. The detection of a large number of FRBs (presently only 
17 FRBs\footnote{https://astro.uni-bonn.de/$\sim$tauris/NS2015/Keane\_FRBs.pdf} are known) is required in order for us to have 
a better understanding of the nature of the population. 
This will allow one to establish whether they form a population of standard candles or some other distribution. Also many 
lines-of-sights will be needed to make them useful probes of the intergalactic medium
and for determining the presence of the so-called missing baryons. With a larger population of sources it will also be possible 
to determine the spectral index of the radio emission which is a key element for probing the emission physics. 
Magnetar flares (\cite{thornton13,kulkarni14}), blitzars \citep{falcke14}, super-giant pulses from neutron stars \citep{cordes15}, and 
pulsar-planet systems \citep{mottez14}, are some of the models suggesting extragalactic origin of FRBs at cosmological distances. 

The known pulsar population is increasing with the surveys running at major single dish and array telescopes 
around the world: e.g. High Time Resolution Universe (HTRU) survey at Parkes \citep{keith10}, HTRU-North survey at Effelsberg \citep{barr11}, 
SUrvey for Pulsars and Extragalactic Radio Bursts (SUPERB) survey\footnote{https://sites.google.com/site/publicsuperb/} at Parkes, Pulsar survey 
Arecibo L-band Feed Array (PALFA, \cite{cordes06}), Green Bank Telescope drift scan survey \citep{boyles13}, Green Bank Northern Celestial Cap (GBNCC) survey \citep{stovall14}, 
 Arecibo all-sky 327 MHz drift pulsar survey (AO327, \cite{deneva13}), the LOFAR Pulsar Pilot Survey (LPPS) survey \citep{coenen14}, the  LOFAR Tied-Array All-Sky 
Survey (LOTAAS)\footnote{http://www.astron.nl/lotaas} at LOFAR and {\it Fermi} directed MSP surveys \citep{ray12}.

Because of the generally steep spectral nature of pulsars, lower frequencies are an obvious choice for searching for fainter 
pulsars away from the Galactic plane, where search sensitivity is not severely affected by sky temparature and increased scattering. Such 
surveys away from the Galactic plane will detect relatively older pulsars. 
The success of {\it Fermi} directed searches in discovering more than 60\% of Fermi MSPs\footnote{http://astro.phys.wvu.edu/GalacticMSPs/GalacticMSPs.txt} 
at frequencies below 1 GHz demonstrates the advantages of low frequency searches \citep{ray12}. 
Besides the GBT and the LOFAR, the Giant Metrewave Radio Telescope (GMRT\footnote{http://gmrt.ncra.tifr.res.in}) is 
another facility which has a sensitive low frequency ($<$ 600 MHz) observing capability. With the aid of reduced quantised
 noise supported by the flexible GMRT Software Backend, (GSB; \cite{roy10}) the search sensitivity to pulsars in general is improved by 30\% 
 compared to its old hardware counterpart. An additional 30\% increase in sensitivity to MSPs is achieved through using the higher time 
resolution mode implemented in the GSB. Thus, effectively, the search sensitivity to normal pulsars is improved by 30\% whereas for MSPs 
a 60\% improved sensitivity is achieved with the GSB \citep{roy13a}. This is exemplified by the discovery of seven MSPs from {\it Fermi} 
directed searches by \cite{bh13}. These were the first Galactic MSPs discovered with the GMRT. This clearly 
illustrates the potential of low frequency pulsar search using the GMRT.

Since we have no information on most pulsar positions a priori (unlike the ones associated with unidentified Fermi sources or supernovae remnants), 
blind searches are required for pulsar discoveries.
 Lower frequency surveys benefit from a larger field-of-view, in addition to the steep spectral nature of pulsars. The success of targeted 
 searches with the GMRT makes it evident that blind searches with the GMRT are warranted and can produce a significant science 
yield in terms of discovering new pulsars, MSPs and transient events. Thus we are carrying out the GMRT High 
Resolution Southern Sky (GHRSS\footnote{http://www.jb.man.ac.uk/research/pulsar/Resources/ghrss.html}) survey for pulsars and 
transients to conduct one of the most sensitive and highest resolution surveys for the southern hemisphere. To make the search 
more efficient and complementary we target a part of the southern sky which has not been searched at frequencies below 1.4 GHz for 
the last two decades. New observing modes were developed to increase the time and frequency resolution of the survey, making it 
even more sensitive to MSPs and normal pulsars with higher DM values. 

Large uncertainties associated with the discovery positions (e.g. $\pm$ 40$'$ for GMRT$-$322, $\pm$ 18$'$ for GBT$-$350, 
$\pm$ 7$'$ for Parkes$-$Lband) hinder sensitive follow up studies of these newly discovered pulsars using coherent 
beams of array telescopes, or at higher frequencies using single dishes. The GMRT interferometric array allows us to 
localise the newly discovered pulsars and 
transients in the image plane with an accuracy of better than $\pm$ 10\arcsec~(half of the typical synthesized beam used 
in the image made at 322 MHz). Precise a priori astrometric positions are also needed to overcome the effect of large 
covariances in the timing fits (with discovery position, pulsar period derivative and unknown binary model in case of 
binary pulsars) for rapid convergence of an initial timing model. Moreover, sensitive coherent array follow up observations 
can significantly reduce the use of telescope time by $\sim$ 20$\times$ for the GMRT (due to the 4 to 5 times sensitivity improvement
for coherent array observations) and also improves the uncertainties in time-of-arrivals (TOAs) or generates more closely 
spaced TOAs in order to avoid ambiguous phase connection while timing. 

Finding pulsars is only half the job. The other half is to understand their nature, rotation properties, spectral indices and relation
to the pulsar population; for this regular follow up is needed. Precise measurements of these parameters puts a new discovery in
its proper place in the pulsar population. Equipped with the precise localisation we carry out more sensitive follow up observations with 
the coherent array of the GMRT. In this paper we demonstrate that convergence in the timing fit is facilitated by a priori knowledge of a precise position.

All the known FRBs detected to date have been found with single dish telescopes, and for most of them the uncertainty in position is 
of order of $\pm$ 7\arcmin~or more. Interferometric imaging using the simultaneously recorded visibility data at 2 s interval will allow 
precise localisation of the FRBs detected from the GHRSS survey. This will allow multi-wavelength follow up and thus the 
possibility of identification of putative host galaxies. However, this will only be feasible for the relatively 
stronger ($>$~3 Jy) (detailed in Section \ref{FRB_loc}) FRBs.  Thus we will be able to 
localise bright FRBs (e.g. Lorimer bursts, \cite{lorimer07}).   

In this paper we describe the survey, the data analysis pipeline, discovery parameters of the 10 pulsars, the localisation of 4 
and timing solution of 2 pulsars discovered in the GHRSS survey.
The observing system and target sky are detailed in Section \ref{sec:obs_sys} and  Section \ref{sec:sky_coverage} of the paper. 
The prediction of survey outcome is detailed in Section \ref{sec:survey_prediction}. The survey processing is discussed in 
Section \ref{sec:processing}. In Section \ref{sec:sensitivity} we calculate the theoretical sensitivity and compare that with 
the sensitivity achieved in the survey. Section \ref{sec:redection} describes the re-detection of known pulsars from this survey and 
their interferometric flux measurements. In Section \ref{sec:discoveries} we give the basic parameters of the pulsars 
discovered in the GHRSS survey and in Section \ref{sec:localisation} we detail the localisation of detected events using 
the GMRT interferometric array. The timing and follow up of the pulsars discovered in the GHRSS survey is described in 
Section \ref{sec:timing}. The relevance of GHRSS survey results and methods with the future Square Kilometre Array (SKA) 
is discussed in Section \ref{sec:SKA}. Section \ref{sec:conclusion} presents the summary of this paper.
  \begin{figure}[!ht]
    \subfloat[\label{fig:512ch}]{%
      \includegraphics[width=3.2in,angle=0]{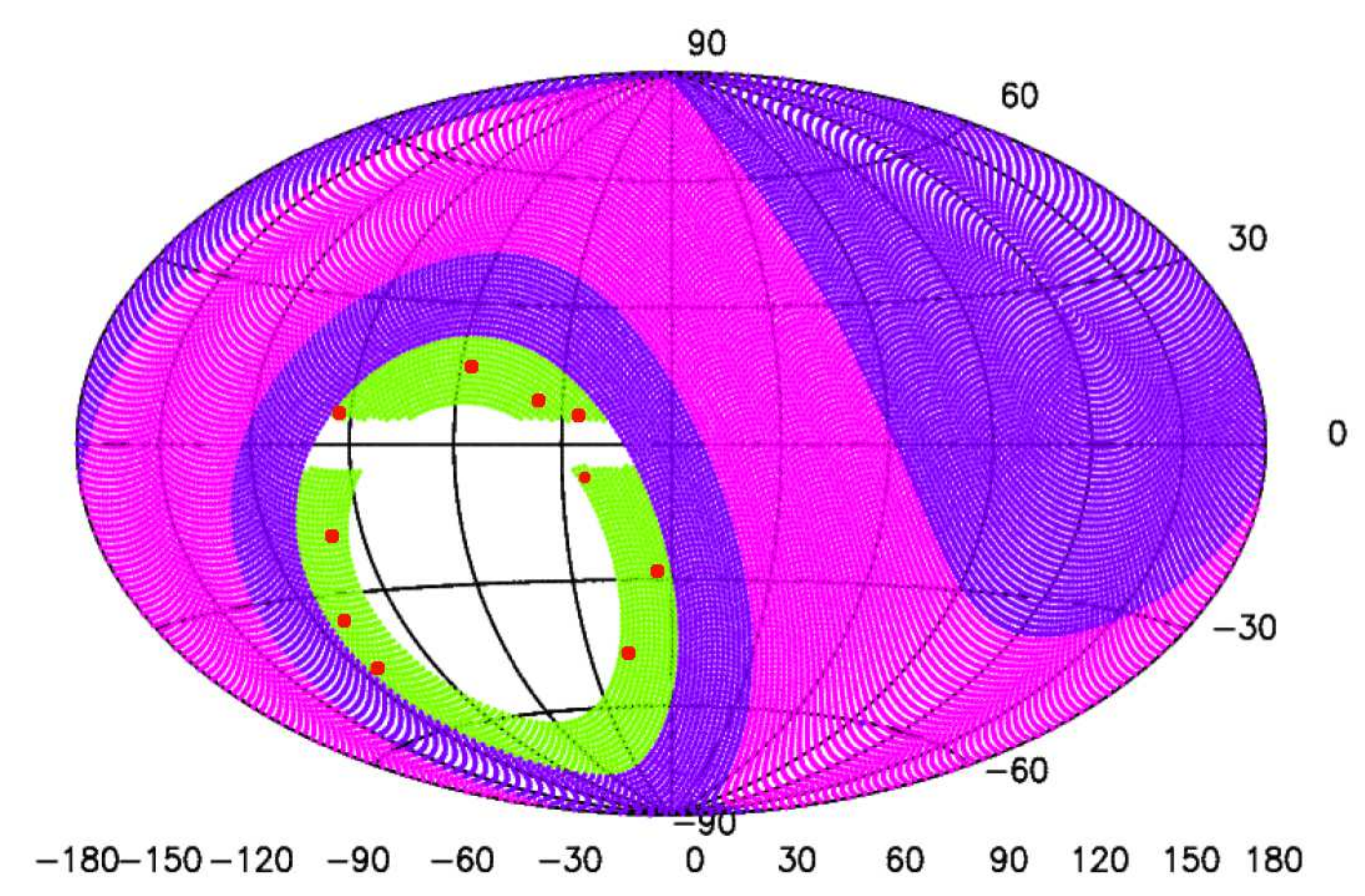}
    }
    \hfill
    \subfloat[\label{fig:2048ch}]{%
      \includegraphics[width=3.2in,angle=0]{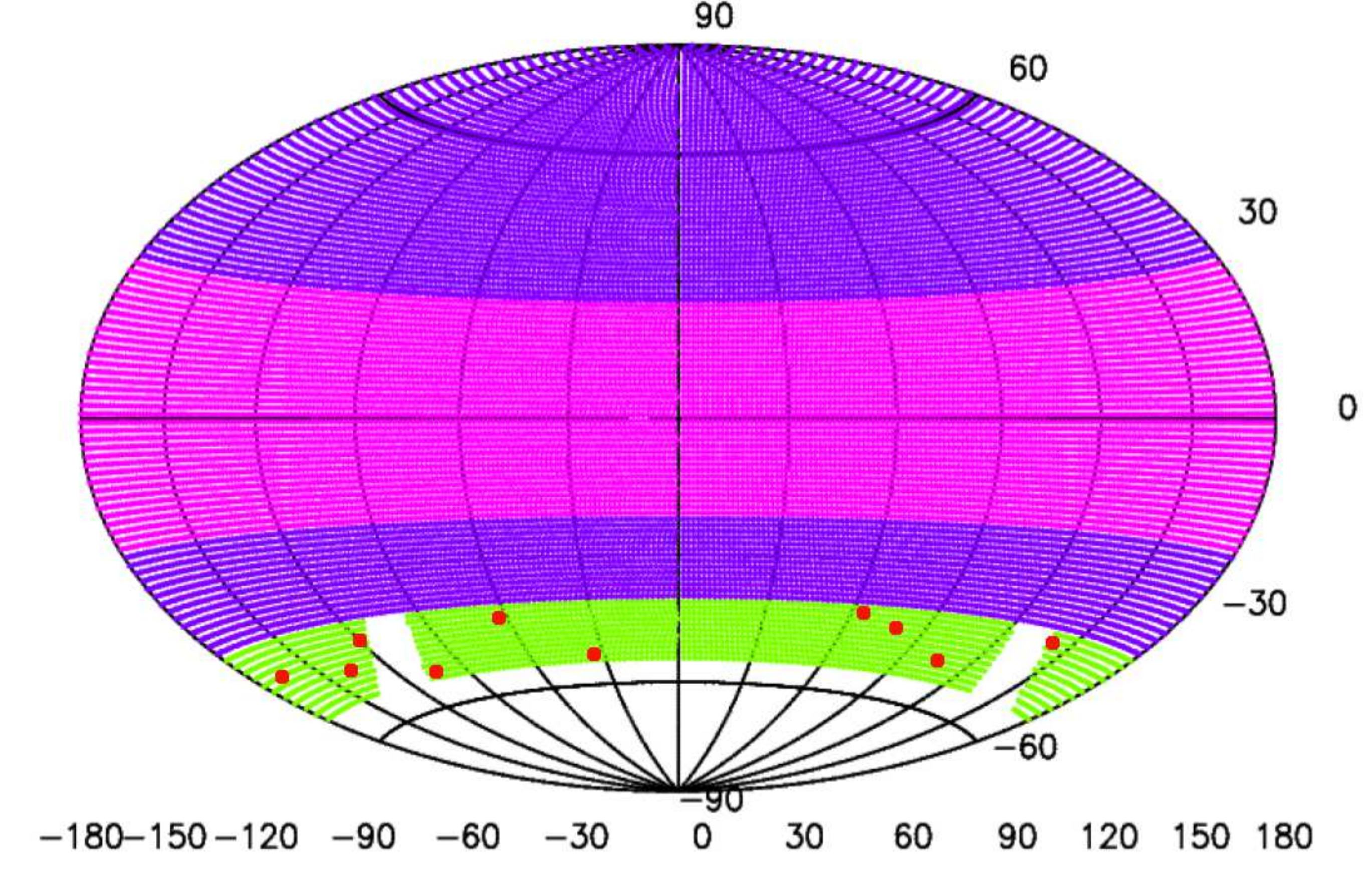}
    }
    \caption{(a) Sky coverage in Galactic coordinates of recent and ongoing pulsar surveys around 300 MHz; GBT drift scan (pink) at 350 MHz, GBNCC (purple) at 350 MHz and GHRSS (green) at 322 MHz\label{fig:sky_coverage}. Red dots are the pulsars discovered with the GHRSS survey. (b) Sky coverage in RA and DEC.}
    \label{fig:sky_coverage}
  \end{figure}
\begin{table*}
\begin{center}
\caption{Survey parameters of the GHRSS survey for the MGL and HGL components}
\vspace{0.3cm}
\label{survey_para}
\begin{tabular}{|l|c|c|c|c|c|c|c|c|c|c|c|c|c|c|c|c}
\hline
Survey         & MGL                    & HGL  \\
               & with HiRes1            & with HiRes2   \\\hline
Galactic region& 5$<$$|b|$$<$20\degr~         & $|b|$~$>$~20\degr\\
Declination    & $-$40\degr$<$Dec$<$$-$54\degr & $-$40\degr$<$Dec$<$$-$54\degr \\
Integration time& 1200 s                & 900 s \\
Sampling time &  61.44 $\mu$s               & 30.72 $\mu$s\\
Bandwidth     &  32 MHz                 & 32 MHz\\
Number of channels & 2048               & 1024\\
Frequency Resolution &15.625 kHz        & 31.25 kHz\\
Number of pointings & 682                & 911 \\
Sky coverage &  1227 deg$^2$            & 1639 deg$^2$\\
Data/pointing & 37 GB                   & 28 GB\\
Total data    & 25 TB                   & 25 TB\\\hline
\end{tabular}
\end{center}
\end{table*}
\section{Observing system}
\label{sec:obs_sys}
The GMRT is a multi-element aperture synthesis telescope consisting of 30 antennas, each of 45 m diameter, spread over a
region of 25 km diameter and operating at 5 different wave bands from 150 MHz to 1450 MHz \citep{swarup97}.
The GMRT Software Backend, built using COTS components, is a fully real-time backend that supports
a FX correlator and a beamformer for an array of 32 dual polarized signals, Nyquist sampled at 33 or 66 MHz \citep{roy10}.
We developed high time and frequency resolution observing modes that are ported to the GMRT Software Backend, producing 
incoherent or coherent beam 2048$\times$0.016275 MHz filter-bank outputs sampled every 61.44 $\mu$s (hereafter HiRes1) 
or 1024$\times$0.03255 MHz sampled every 30.72 $\mu$s (hereafter HiRes2). This translates to a data rate of 32 MB/s for 
the high time-resolution data with 8 bits/sample after adding two polarized intensities. In parallel 
we recorded frequency averaged 512 channel visibilities for each of the 465 baselines at 2 s intervals resulting in a 2 MB/s data rate. 
The target scans were interleaved with scans on nearby calibrator sources every 1 hour to allow for phase calibration for imaging. 
The visibility data were then converted to FITs files for further processing in {\sc aips}\footnote{http://www.aips.nrao.edu/index.shtml}/{\sc casa}\footnote{http://casa.nrao.edu}. 
The GMRT interferometric array provides the opportunity for combining the imaging mode with beamforming to widen the scope of studying pulsars (e.g. \cite{roy12,bh13,roy13}). 
For the GHRSS survey we used the incoherent array beamformer output in order to maximise the field-of-view. In order to 
optimise the gain of the incoherent beam, the antenna based gain offsets were calibrated out prior to the addition of the signals.

Combining contributions from receiver temperature, ground temperature and spillover from the ground, the total system temperature
at 322 MHz is 106 K on cold sky\footnote{http://www.gmrt.ncra.tifr.res.in/gmrt$\_$hpage/Users/doc/GMRT$-$specs.pdf}. The data were recorded to
local disks and then transferred to the dedicated 64 TB GHRSS storage at the National Centre for Radio Astrophysics (NCRA) using an 8 Gbps 
fibrelink. We shipped the data disks from the GHRSS storage to the University of Manchester for processing on the {\it Hydrus} supercomputer. 
The {\it Hydrus} supercomputer with a dedicated 64 TB storage was used as a major processing and data-centre host for the GHRSS survey. Raw 
data were archived on LTO5 magnetic tapes immediately after the observations for long term storage. 

\begin{table*}
\begin{center}
\caption{Major ongoing or recently completed off-Galactic plane surveys}
\label{survey_comp}
\vspace{0.3cm}
\begin{tabular}{|c|c|c|c|c|c|c|c|c|c|c|c|c|c|c|c|c|}
\hline
Survey name               & Frequency           & Sky coverage                & Discoveries       & Sensitivity$^\dagger$\\
$-$ Telescope             & of search           &                             &                 &                    \\
                          &   (MHz)             &                             &                 &   (mJy)           \\\hline
HTRU$^1$                  & 1352                & $-$120\degr$<$l, l$<$30\degr    & 104 PSR, 26 MSP &  1.5                  \\
                          &                     & $|b|$$<$15\degr                &                 &                    \\
$-$ Parkes                &                     & 4500 deg$^2$                &                 &                      \\
HTRU$-$N                  & 1360                & $|b|$~$>$~15\degr, Dec~$>$~$-$20\degr & 12 PSR          &  1.5                  \\
$-$ Effelsberg            &                     &                             &                 &                      \\
GBNCC$^2$                 & 350                 & Dec$>$$-$40\degr                & 108 PSR, 12 MSP &  0.6                 \\
$-$ GBT                   &                     & 19500 deg$^2$               &                 &                     \\
GBTdriftscan$^3$          & 350                 & $-$21\degr~$<$Dec$~<$26\degr      & 26 PSR, 7 MSP   &  0.9                \\
$-$ GBT                   &                     &                             &                 &                   \\
AO327$^4$                 & 327                 & 0\degr~$<$Dec$<$~28\degr        & 24 PSR, 3 MSP   &  0.3                 \\
$-$ Arecibo               &                     &                             &                 &                     \\
LOTAAS$^5$                & 135                 & Dec~$>$0~\degr                 & 18 PSR          &  0.3                 \\
$-$LOFAR                  &                     & 5156 deg$^2$                &                 &                    \\
GHRSS$^\ddagger$$^6$      & 322                 & $-$20\degr~$<$Dec~$<$~$-$54\degr & 10 PSR, 1 MSP   &  0.5              \\
$-$ GMRT                  &                     & 1000 deg$^2$                &                 &                    \\\hline
\end{tabular}
\end{center}
{$\dagger$: Sensitivity calculated at 322 MHz considering 5$\sigma$ limit for 10\% duty cycle for spectral index of $-$1.7\\
$\ddagger$: Full GHRSS survey will cover 2866 deg$^2$.\\
$^1$: http://researchbank.swinburne.edu.au/vital/access/manager/Repository/swin:31985\\
$^2$: http://arcc.phys.utb.edu/gbncc/ \\
$^3$: http://astro.phys.wvu.edu/GBTdrift350/\\
$^4$: http://www.naic.edu/\~deneva/drift-search/\\
$^5$: http://www.astron.nl/lotaas\\
$^6$:  http://www.jb.man.ac.uk/research/pulsar/Resources/ghrss.html}
\end{table*}
\section{Sky coverage}
\label{sec:sky_coverage}
The 322 MHz GHRSS survey consists of scans of the sky (ranging from $-$40\degr~ to $-$54\degr~in declination), 
complementary to other low frequency surveys at the GBT, and LOFAR. 
Figure \ref{fig:sky_coverage} shows portion of the target sky (marked in green) compared to the GBT surveys at 350 MHz, 
noting that LOFAR is not scanning below declination zero. Though the pink and purple bands are indicative of GBT drift scan 
survey and GBNCC surveys, there are empty patches in the drift scan survey, which are being covered by the GBNCC survey.

The target sky for the GHRSS survey has not been searched at frequencies below 1.4 GHz for the last two decades. 
The Parkes southern sky survey at 436 MHz with $\sim$3 mJy search sensitivity \citep{bailes94} was conducted from 1991 to 
about 1995, and it discovered 85 normal pulsars and 17 MSPs \footnote{http://www.jb.man.ac.uk/\~pulsar/research/jodsum/node3.html\#s70}. 
 
For the GHRSS survey we are not considering low Galactic latitudes ($|b|$$<$5\degr). The target sky has two components, 
a mid-latitude component, MGL, for 5$<$$|b|$$<$20\degr~covering the sky with declination range $-$40\degr~to $-$54\degr~with 
1200 s integration with HiRes1 mode, and a high latitude component, HGL, ($|b|$$> $20\degr) covering 
the same declination range with 900 s integration with HiRes2 mode. 
The range of sky temperature for the GHRSS survey range from 33 K to 220 K based on the \cite{haslam82} 408 MHz map. 

In order to partly compensate for the $\sim$ 50\% increase of 
sky temperature at MGL ($|b|$$\sim$ 10\degr) with respect to HGL, we choose to increase the integration time to 1200 s.
The target sky is covered with 682 and 911 pointings respectively for the MGL and HGL components, totaling 1953 pointing scans, 
each covering 1.8 deg$^2$ adding up to a total coverage of 2866 deg$^2$. The survey parameters are summarised in Table \ref{survey_para}.
\begin{figure}[htb]
\begin{center}
\includegraphics[width=4.5in,angle=0]{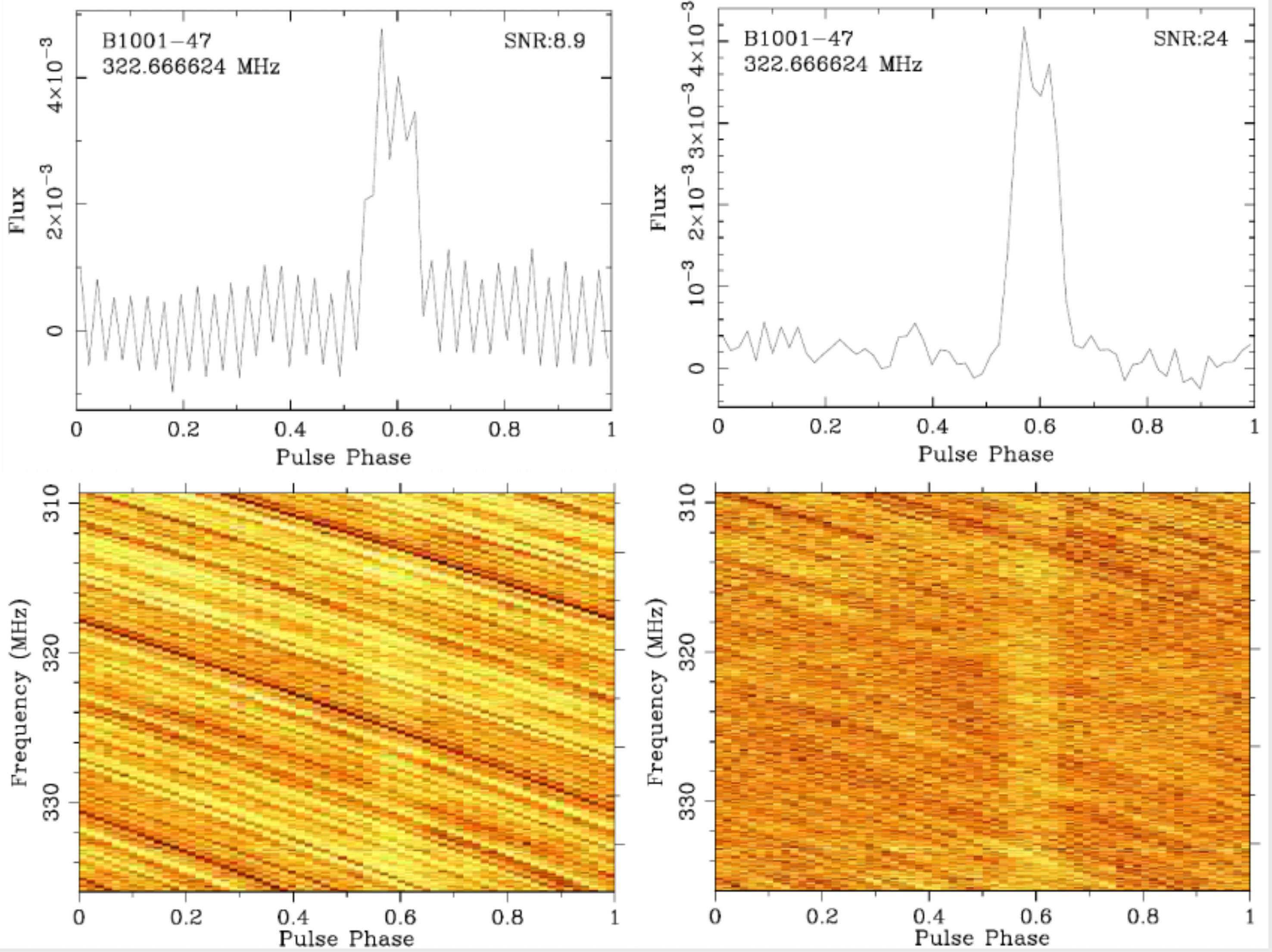}
\caption{Integrated profile and phasogram of PSR B1007$-$47 for before (left panel) and after (right panel) zero-DM RFI mitigation. This
observation was unusually affected by RFI. Even with some residual RFI present, a significant cleaning of the data and improvement of SNR by
about a factor of 3 is achieved. The zero-DM RFI mitigation is followed by further cleaning of the data using the {\sc rfifind} package of {\sc presto}
resulting in 8\%-10\% improvement.}
\label{fig_rfi}
\end{center}
\end{figure}
\begin{figure}[htb]
\begin{center}
\includegraphics[width=4.5in,angle=0]{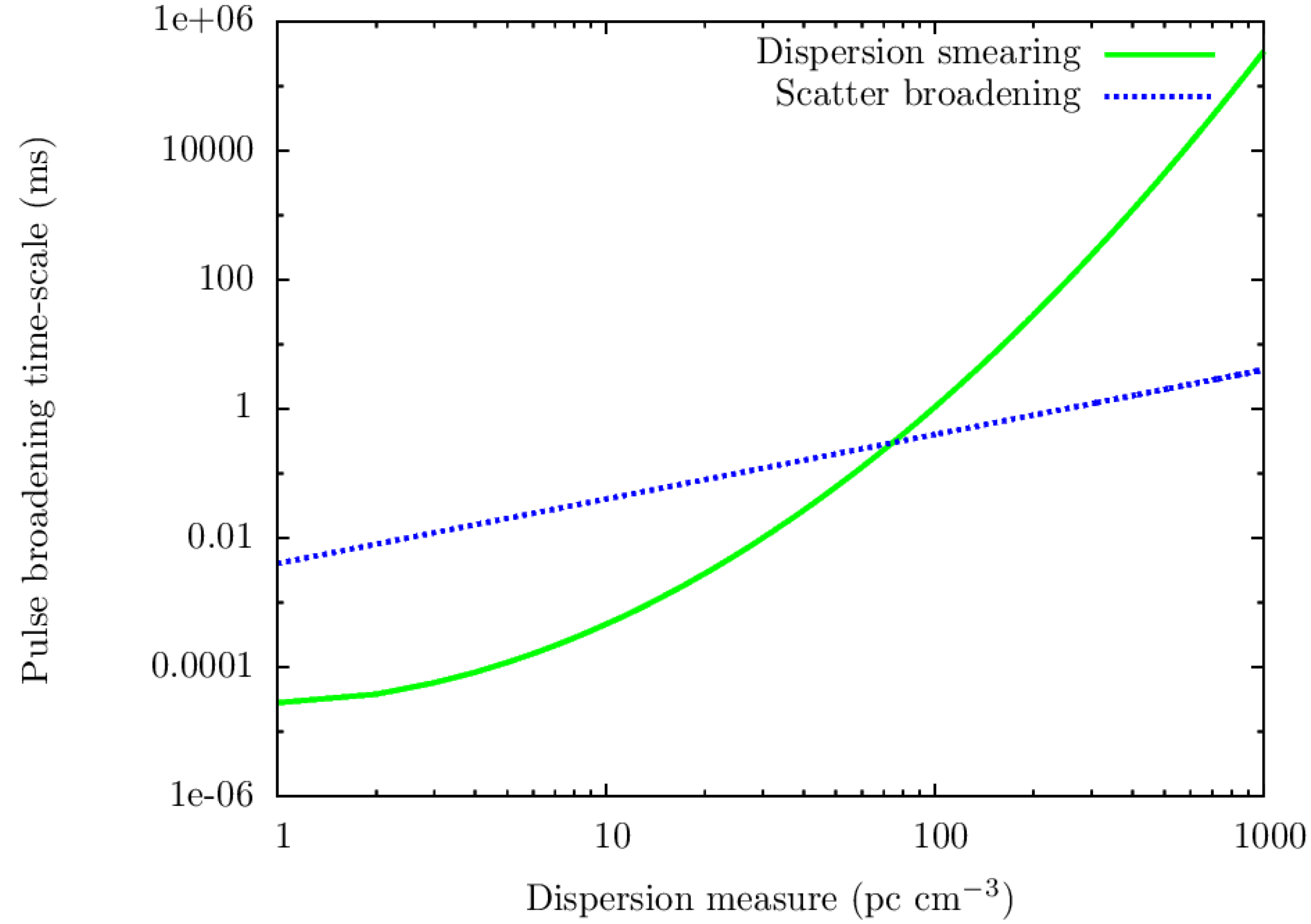}
\caption{Scatter broadening and dispersion smearing as a function of DM. At DM$<$75 pc~cm$^{-3}$ scatter broadening dominates over dispersion smearing.
Thus the GHRSS survey sensitivity is not significantly affected by the instrumental broadening up to DM of 75 pc~cm$^{-3}$.}
\label{fig_DM_scatter}
\end{center}
\end{figure}
\begin{figure}
\centering
\begin{tabular}{c c}
\subfloat{\includegraphics[scale=0.4]{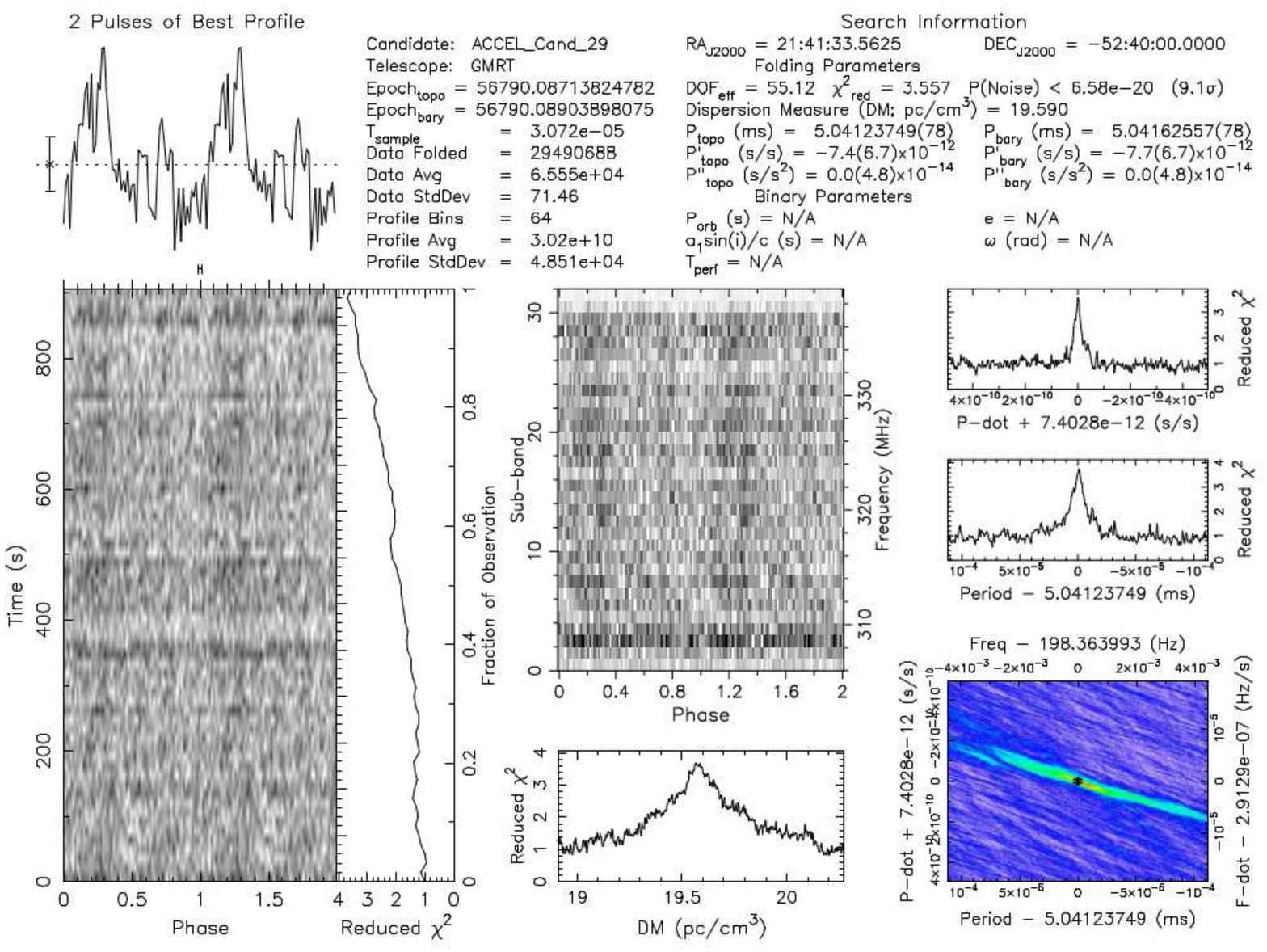}}
\\
\multicolumn{2}{c}
\subfloat{\includegraphics[scale=0.4]{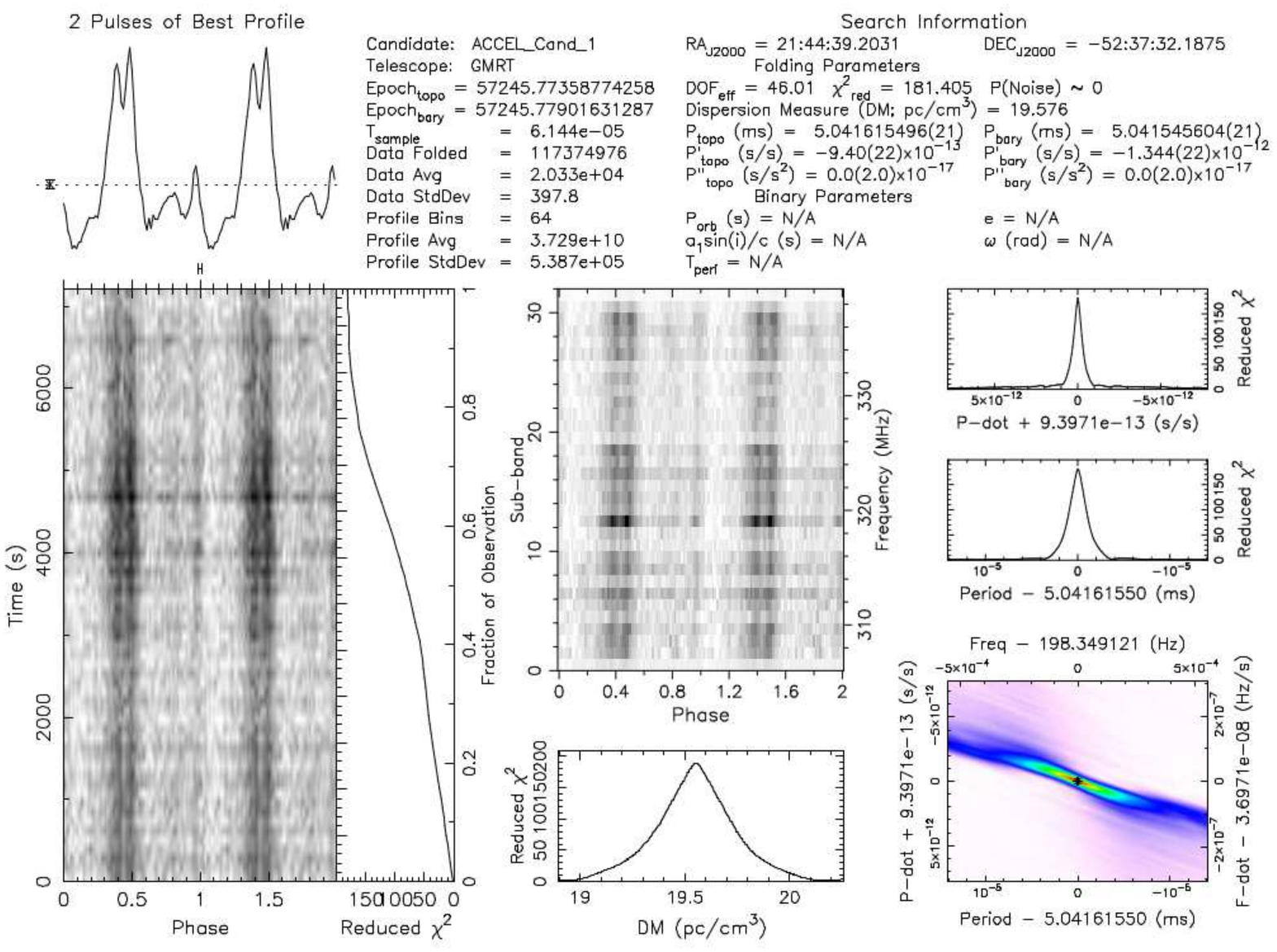}}
\end{tabular}\caption{ Top panel : Example diagnostic plot from the GHRSS periodicity search pipeline. This plot shows the MSP J2144$-$5237 discovered in 
the GHRSS survey with incoherent array observations, Bottom panel: {\sc presto} search output of PSR J2144$-$5237 with a coherent array observation. The plot 
demonstrates the improved detection significance with coherent array observations which can be performed after precise localisation of the pulsar in the 
image plane (Section \ref{sec:localisation}).}
\label{fig_incoh_coh}
\end{figure}
\section{Survey prediction}
\label{sec:survey_prediction}
In order to determine the discovery potential of the GHRSS survey we performed simulations based on 
the {\sc psrpoppy}\footnote{https://github.com/samb8s/PsrPopPy}code \citep{bates14}. {\sc psrpoppy} is a modified 
version of {\sc psrpop} \citep{lorimer06} that models the period, luminosity and spatial distributions of a set 
of pulsars discovered in the successful Parkes Multibeam Pulsar Survey \citep{manchester01} and 
the high-latitude pulsar survey \citep{burgay06} to predict the discovery potential of the ongoing 
surveys (e.g. \cite{keith10,keane15}). For the purpose of simulation we considered normal pulsars defined as $P>$~30 ms, $\dot{P}<$ 1$\times$ 10$^{-12}$. 
In simulation of the expected normal pulsar population from the GHRSS survey, 
we used the input parameters listed in Table \ref{survey_simulations} (which are generally default options of {\sc psrpoppy} and are 
similar to the ones used in \cite{keith10}).\\
\begin{table*}
\begin{center}
\caption{Parameters for {\sc psrpoppy} simulations}
\vspace{0.3cm}
\label{survey_simulations}
\begin{tabular}{|c|c|c|c|c|c|c|c|c|c|c|c|c|c|c|c|}
\hline
Parameters                                           & Normal  pulsars                                            & Millisecond pulsars \\
                                                     &                                                            &                                \\\hline
Spin Period  Distribution                            &  Log-normal                                                & \cite{lorimer12}$^\dagger$          \\
$log_{10}\langle$P (ms)$\rangle$                                  &  2.7                                                       &                                \\
std $log_{10}\langle$P (ms)$\rangle$                               & 0.34                                                       &                                \\
Luminosity Distribution                              & Log-normal                                                 &     Power-law           \\
                                                     & $\langle$$log_{10}$ L (mJy kpc$^2$ )$\rangle=$ $-$1.1           &    L(mJy kpc$^2$)$_{min}=$  0.1   \\
                                                     & std($log_{10}$ L (mJy kpc$^2$ ))$=$ 0.9                         &    L(mJy kpc$^2$)$_{max}=$ 1000      \\
                                                              &                                                                                     &    pow=$-$1.45\\
Galactic Scale Height                            &  Exponential                                                                &    Exponential                      \\
                                                              &  0.33 kpc                                                                      &     0.5 kpc \citep{levin13}                                       \\
Radial Distribution  model                     &   \cite{lorimer06}                                                          & \cite{lorimer06} \\
Galactic electron distribution                 & NE2001 model                                     &  NE2001 model       \\
                                               & \citep{cordes02}                                 &  \citep{cordes02}       \\
Spectral index distribution                     & Gaussian                                                                    &  Gaussian                                                  \\
$\langle$$\alpha$$\rangle$                   & $-$1.7                                                                         &  $-$1.7                                           \\
std ($\alpha$)                                        & 0.35                                                                            &  0.35                                                      \\\hline
\end{tabular}
\end{center}
$\dagger$: with this built-in option in {\sc psrpopoy}, periods of millisecond pulsars are picked in at random from data base provided in \cite{lorimer12}.
\end{table*}
Using the parameters listed in Table \ref{survey_simulations} and considering a total of 30,000 possibly detectable pulsars, the {\sc psprpoppy} 
software generates a model \citep{bates14} such that with Parkes Multibeam Pulsar Survey, 1038 pulsars are detected, which is close to the 1122 pulsars 
actually discovered in this survey. 
With this model, we ran the simulation 100 times with the specifications of the GHRSS survey and predict that it would discover 97$\pm$7 
pulsars, considering spectral index mean of $-$1.7. Simulations considering different spectral indices with mean of $-$1.4 and $-$1.8 result in 
discovery predictions of about 81$\pm$6 and 113$\pm$8 normal pulsars respectively. 
\begin{figure}[htb]
\begin{center}
\includegraphics[width=4in,angle=0]{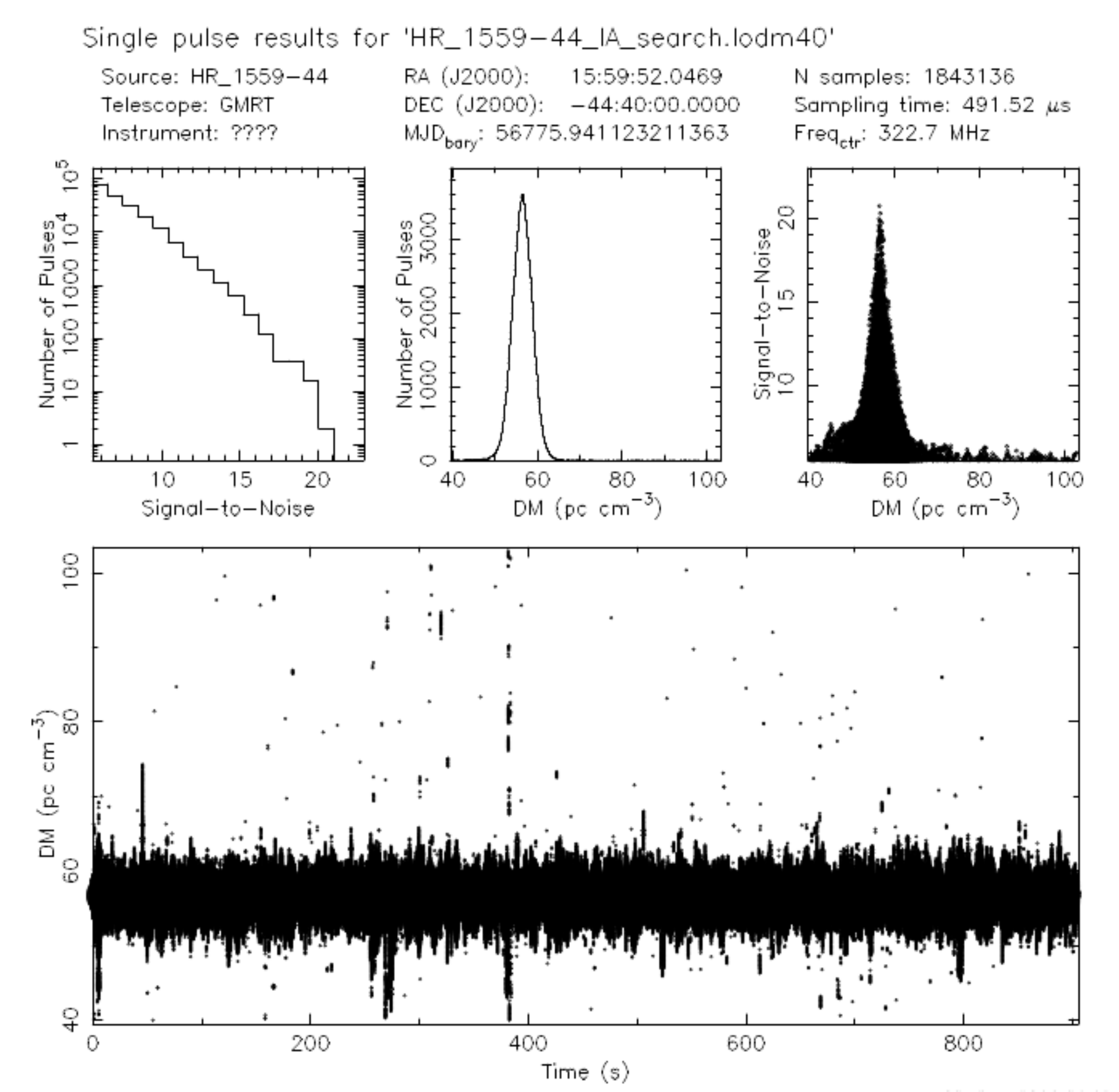}
\caption{Example output of our {\sc presto} based single-pulse search pipeline using CPU processing, showing bright single pulses from PSR J1559$-$4438 (which is a 
known pulsar re-detected in the GHRSS survey) around a DM of 56 pc~cm$^{-3}$.
The top left panel shows the signal-to-noise distribution of the detected pulses, number of pulses versus trial DM (in top centre), and signal-to-noise as
a function of trial DM. The lower plot shows signal-to-noise of events as a function of time and trial DM.}
\label{fig:single_cpu}
\end{center}
\end{figure}
We expect to find considerably fewer MSPs due to their intrinsic faint nature in addition to dispersion smearing and scatter broadening.
For the simulation of the MSP population we considered a modified input for the spin period distribution, height above the Galactic plane and 
luminosity distribution as listed in Table \ref{survey_simulations}. According to the {\sc psrpoppy} simulation the GHRSS survey will 
discover around 9$\pm$3 MSPs considering spectral index mean of $-$1.7. Simulations considering spectral indices with mean of $-$1.4 and $-$1.8 resulted 
in discovery predictions of about 7$\pm$3 and 11$\pm$4 MSPs respectively.    
\begin{figure}[htb]
\begin{center}
\includegraphics[width=5in,angle=0]{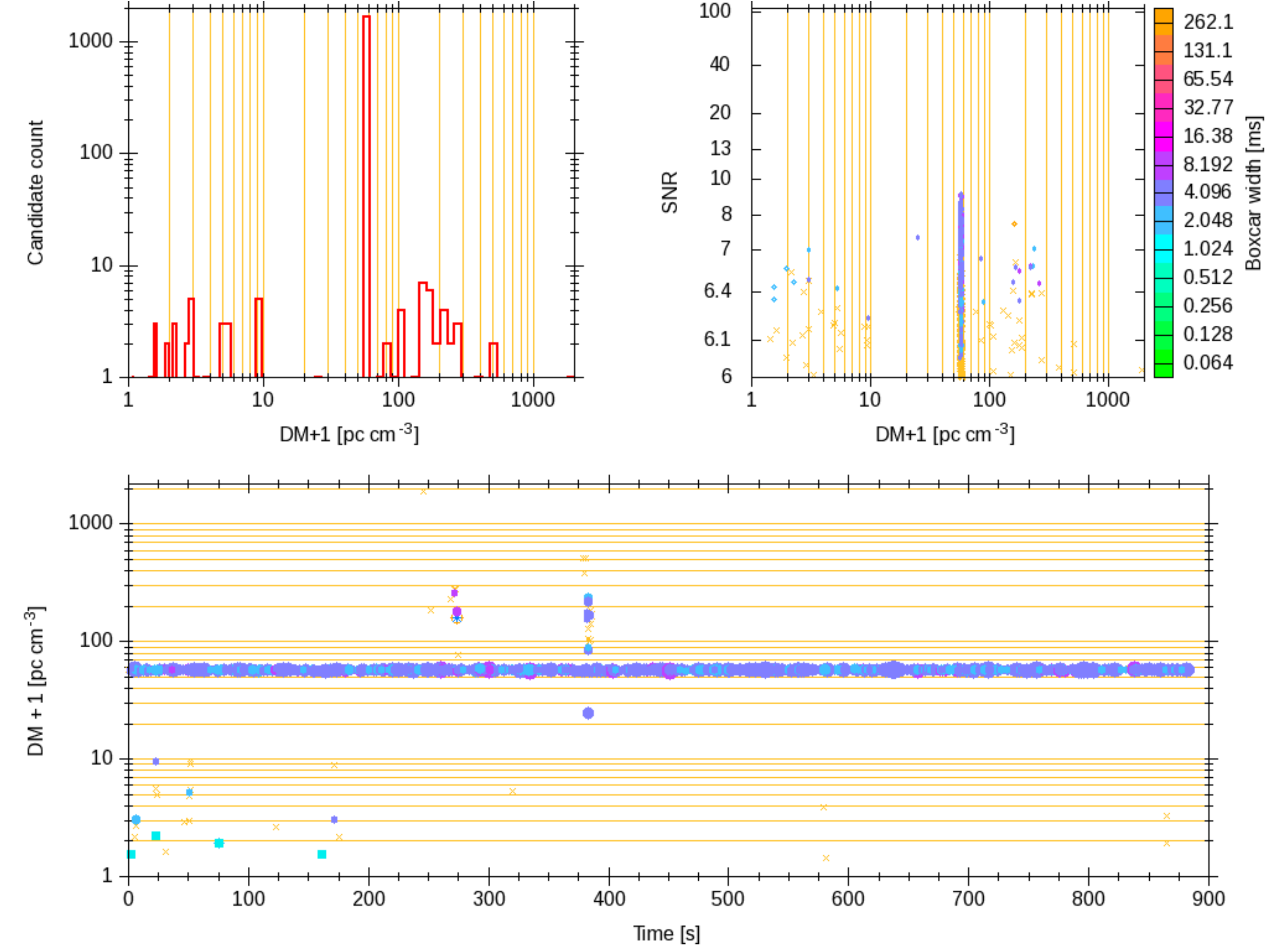}
\caption{Example output of {\sc biforst} single pulse search pipeline with GPU on the known pulsar J1559$-$4438 (re-detected in the GHRSS survey) 
showing bright pulses around a DM of 56 pc~cm$^{-3}$. The top left panel shows the detected pulses versus DM and top right panel shows signal-to-noise 
as a function of trial DM. The lower plot shows signal-to-noise of 
events as a function of time and trial DM.}
\label{fig:single_gpu}
\end{center}
\end{figure}

We also tried to judge the discovery potential of the GHRSS survey by comparing with the ongoing 
GBNCC survey which is scanning the sky at a similar frequency and with comparable sensitivity to the GHRSS survey i.e. about 0.5 mJy of 
minimum detectable flux density (detailed in Section \ref{sec:sensitivity}).
GBNCC survey has covered 19500 deg$^2$ and has discovered 108 pulsars\footnote{http://arcc.phys.utb.edu/gbncc/}. 
Scaling from the GBNCC areal discovery rate, we expect to find of the order of 16 pulsars in the total GHRSS sky coverage (2866 deg$^2$).
 
In addition to the potential of discovering pulsars and millisecond pulsars, the GHRSS survey provides a snapshot of the 
transient radio sky in the southern hemisphere at 322 MHz. Based on the 35\% completion of HTRU-South survey analysis, 
\cite{thornton13} estimated the FRB rate as 1$^{+0.6}_{-0.5}$$\times$10$^4~$sky$^{-1}~$day$^{-1}$ for bursts having 
fluence $>$~3 Jy ms. For the GHRSS survey we calculated a minimum detectable flux of 1.6 Jy for 10$\sigma$ detection 
with 5 ms bursts (detailed in Section \ref{sec:sensitivity}). Coincidentally this minimum detectable flux for an FRB 
translates to a fluence of $\sim$ 3 Jy ms. Considering the total 2866 deg$^2$ of the GHRSS survey area with $\sim$ 15 mins 
of dwell time, we expect to discover 7$^{+5}_{-3}$ FRBs having fluence $>$ 3 Jy ms. However this prediction is based on 
the assumption that FRBs have a flat spectrum all the way from 1.4 GHz to 300 MHz and of course scattering and spectral 
index are largely unknowns for FRBs. Thus, the predictions are likely be lower than these estimates unless they are steep 
spectrum objects, and scattering at low frequencies is not large.
Moreover the uncertainty on the rate presented by \cite{thornton13} and new lower rate estimates from \cite{rane15}, 
\cite{champion15} and \cite{petroff14}, the number of detected FRBs could be a factor of a few lower. However, any GHRSS 
detections will help us to significantly improve rate determinations and the frequency dependence of FRBs.

The comparison of survey predictions with achieved discovery rate of pulsars and transients from the GHRSS survey is discussed 
in Section \ref{sec:discoveries}.

\section{Survey processing}
\label{sec:processing}
To analyse the survey data, we used the 1456 core {\it Hydrus} cluster (30 Tflops) of the Jodrell Bank pulsar group 
and 512 core {\it IBM} cluster (10 Tflops) at the NCRA. The full GHRSS 
survey will generate approximately 50 TB of data and so far we have collected about 17 TB data corresponding to 35\% 
completion of the survey. To optimise the data processing (i.e. considering the data transfer time from the 
observatory to the University of Manchester) and maximise the usage of the available CPU cycles on the available 
computing clusters we analysed 50\% of the data with {\it Hydrus} and the rest with the {\it IBM} cluster.
\subsection{RFI flagging} 
Like most low frequency radio telescopes, the GMRT is affected by local RFI. Frequently observed RFI occurrences at the GMRT include 
broadband impulsive signals originated from spark discharges from power-lines, narrow spectral impurity and short time-domain bursts. 
We performed zero-DM subtraction (Eatough et al. 2009) to remove broad band interference. The zero-DM subtraction procedure 
significantly reduced the number of spurious candidates detected in periodicity searches and in single pulse searches. This 
dramatically improved the detection significance towards broad band impulsive signals coming 
from pulsars. Figure \ref{fig_rfi} is the integrated profile and phasogram of PSR B1007$-$47 before and after zero-DM subtraction. 
Significant RFI can be seen as diagonal stripes caused by pulsed emission from power-lines
or its subsequent harmonics, which are mitigated to a large extent after zero-DM subtraction and significant 
improvement of the pulsed detection is observed. Some residual RFI is still present in the data which can be seen in  
Figure \ref{fig_rfi}. {\sc rfifind} generally cleans up the data, especially for near to zero-DM residual RFIs, narrow-band spectral features and 
short time-domain bursts.  
 We made this zero-DM subtraction routine a part of the publicly available {\sc sigproc} distribution. Fourier domain analysis of the 
time series dedispersed at a DM $= 0$ enabled us to excise the powerline frequency, 50 Hz.
Finally, we employed {\sc presto{\footnote{http://www.cv.nrao.edu/$\sim$sransom/presto/}}}-based {\sc rfifind} package 
on the zero-DM subtracted time series to remove any remaining bursts of interference or strong spectral features. It is noteworthy that 
without these RFI flagging steps, 90\% of the discovered pulsars reported in this paper would not have been detected.\\

\subsection{Dedispersion} 
\label{sec:processing_dedisp}
We dedispersed the data over a range of DMs, 0 to 500 pc~cm$^{-3}$, resulting in about 6000 dispersion trials for HGL and 10000 for MGL. 
The dispersion plan is generated following the {\sc ddplan} of {\sc presto} in order to minimise the loss of sensitivity caused 
by dispersion smearing over a single frequency channel (Table \ref{dispersion_plan}). We reduced the time resolution for higher dispersion measure trials by 
down-sampling the data. For the GHRSS observing mode with 2048 frequency channels, the dispersive smearing within a frequency channel 
for a DM of 500 pc~cm$^{-3}$ is 1.9 ms at 322 MHz. Figure \ref{fig_DM_scatter} shows the scatter broadening \citep{bhat04} 
and dispersion smearing for the GHRSS survey as a function of DM. This indicates that with the newly implemented high resolution mode, the GHRSS survey 
sensitivity is not significantly affected by the instrumental broadening resulting from dispersion smearing up to a DM of 75 pc~cm$^{-3}$.
\begin{table*}
\begin{center}
\caption{Dispersion plan and corresponding time resolution for HGL and MGL according to the {\sc ddplan} of {\sc presto} }
\vspace{0.3cm}
\label{dispersion_plan}
\begin{tabular}{|l|c|c|c|c|c|c|c|c|c|c|c|c|c|c|c|c}
\hline
DM range     & \multicolumn{2}{|c|}{HGL} & \multicolumn{2}{|c|}{MGL} \\
             & DM step (pc~cm$^{-3}$)& Sampling resolution ($\mu$s) & DM step & Sampling resolution ($\mu$s) \\\hline
1-25         &   0.010  &      30.72       & 0.010          &               61.44\\
25-50        &   0.025  &      61.44       & 0.010          &               61.44\\
50-100       &   0.050  &      122.88      & 0.025          &               61.44\\
100-150      &   0.100  &      245.76      & 0.050          &               122.88\\
150-300      &   0.200  &      491.52      & 0.100          &               245.76\\
300-500      &   0.500  &      491.52      & 0.200          &               491.52\\\hline
\end{tabular}
\end{center}    
\end{table*}   
\subsection{Periodicity search} 
The dedispersed time series were Fourier transformed and candidates with periodicities of known RFI sources were removed.
We used a Fourier-based acceleration search method using the standard pulsar search techniques implemented in {\sc presto}. 
The acceleration search spanned a broadening of the Fourier peaks by up to 50 Fourier frequency bins (i.e. $\it z_{max}$ of 50). 
This translates to 5 m s$^{-2}$ line-of-sight acceleration for a 2 ms pulsar over 15 mins of observing scan detected in all 8 harmonics.
The search for periodicity was done using harmonic summing (up to 8 harmonics). The candidates for each DM were stored for comparison.\\

\subsection{Candidate sorting}
Candidates within a Fourier bin of another candidate were considered to be the same. We also ignored the candidates with summed incoherent power less than 4$\sigma$.
Harmonically related candidates were removed, keeping the one with highest detection significance. We considered only those candidates that 
appear in at least three consecutive DMs to trace the DM dependence of signal-to-noise (SNR). A list of candidates was prepared after sorting. 

\subsection{Candidate folding}
The zero-DM subtracted filter-bank data were dedispersed and folded with the periodicity parameters of sorted candidates in 32 sub-integrations and 64 frequency 
bands using {\sc presto} tool {\sc prepfold}. 
The left panel of Figure \ref{fig_incoh_coh} shows an example of typical diagnostic plot from candidate folding for an MSP discovered in the GHRSS survey.

\subsection{Candidate investigation}
A survey such as the GHRSS produces a large number of pulsar candidates, and it takes a lot of human effort to visually inspect the candidates in
a reasonable amount of time. For example, a typical GHRSS survey scan with good RFI conditions produces around 100 to 200 folded candidates which are then 
manually investigated for possible pulsed emission. However, in the case of a moderately bad RFI environment, the number of candidates can be a few thousand 
for each pointing. We employed a neural network based binary classifier capable of separating candidates arising from noise or RFI, from those generated
by radio pulsars following \cite{lyon15}. We used this scoring method and machine learning algorithm to filter the best candidates for further inspection.
The classifier is a very-fast-decision-tree and it uses the statistics (mean, standard deviation, kurtosis and skew) of the candidate profile and how the 
signal-to-noise varies as a function of DM as input scores. The classifier is applied to all periodicity candidates to label those as positive or 
negative, and only the positive candidates are viewed by eye. For example, 7700 candidates from one epoch were reduced to just 156 positive predictions, 
resulting in about 50 times reduction in the number of candidates to be viewed. Out of these positive predictions, 17 were re-detections of known pulsars or their 
harmonics, and 5 were promising pulsar candidates for follow up. We also examined all of the same 7700 candidates by eye to confirm that the classifier 
had performed accurately. The current version of the classifier was trained using a LOFAR data set but will be re-trained using the GHRSS data. The training 
set will include re-detections of known pulsars and discoveries from the GHRSS survey, as well as a sample of noise and RFI candidates from this survey. This 
will improve the performance of the classifier and reduce the number of false positives.

\subsection{Single pulse processing}
Besides the Fourier based periodicity searches, we performed a single pulse search to detect transient events, such as the RRATs or 
FRBs with possible extragalactic origin. We developed a {\sc presto}-based single pulse search pipeline for the {\it Hydrus} 
cluster of the Jodrell Bank pulsar group and {\it IBM} cluster at the NCRA for analysing the GHRSS survey data. After 
filter-bank conversion, zero-DM subtraction and RFI flagging, we dedispersed the raw data file at various trial DM values up to a 
DM of 2000 pc~cm$^{-3}$ following {\sc ddplan}. In order to accelerate the single pulse search we reduced the nominal time resolution to 
1 ms and the dedispersed time series is further down sampled according to {\sc ddplan}. We then searched for single pulses in the 
time series using the {\sc presto} code {\it single\_pulse\_search.py} \citep{McLaughlin06} above a detection significance of 5$\sigma$. 
Then we investigated the diagnostic plots. Figure \ref{fig:single_cpu} shows output from our CPU based single pulse search pipeline for the known pulsar J1559$-$4438. For a detected 
dispersed event we further investigate the dynamic spectrum to confirm the expected quadratic time delay nature of an astrophysical signal.  

\subsection{GPU-based processing}
Since analysis of our data were limited by available compute power, re-processing the data in the future as we upgrade our compute power
will possibly result in additional discoveries.
We are in the process of setting up a multi-node Graphics Processing Unit (GPU) cluster, consisting of 5-nodes each with 4 Nvidia GTX 980 
GPUs\footnote{http://www.geforce.co.uk/hardware/desktop-gpus/geforce-gtx-980/specifications} each, at the University of Manchester to aid 
the GHRSS survey. Re-analysis of the GHRSS data in search for highly accelerated pulsars over a linear drift range of $\pm$ 250 
frequency bins is being performed with a variant of the {\sc peasoup}\footnote{https://github.com/ewanbarr/peasoup} software implemented by us for the GHRSS survey. In 
addition, a GPU based single pulse search pipeline based on {\sc heimdall} \citep{barsdell12} software was optimised and implemented for 
the GHRSS survey. This integrated multi GPU pipeline for pulsar and transient search implemented for the GHRSS survey is 
called {\sc biforst} (paper in preparation). Figure \ref{fig:single_gpu} shows the output from this pipeline for single pulse searches. 

\section{Survey sensitivity}
\label{sec:sensitivity}
  \begin{figure}[!ht]
    \subfloat[\label{fig_sensitivity}]{%
      \includegraphics[width=3.2in,angle=0]{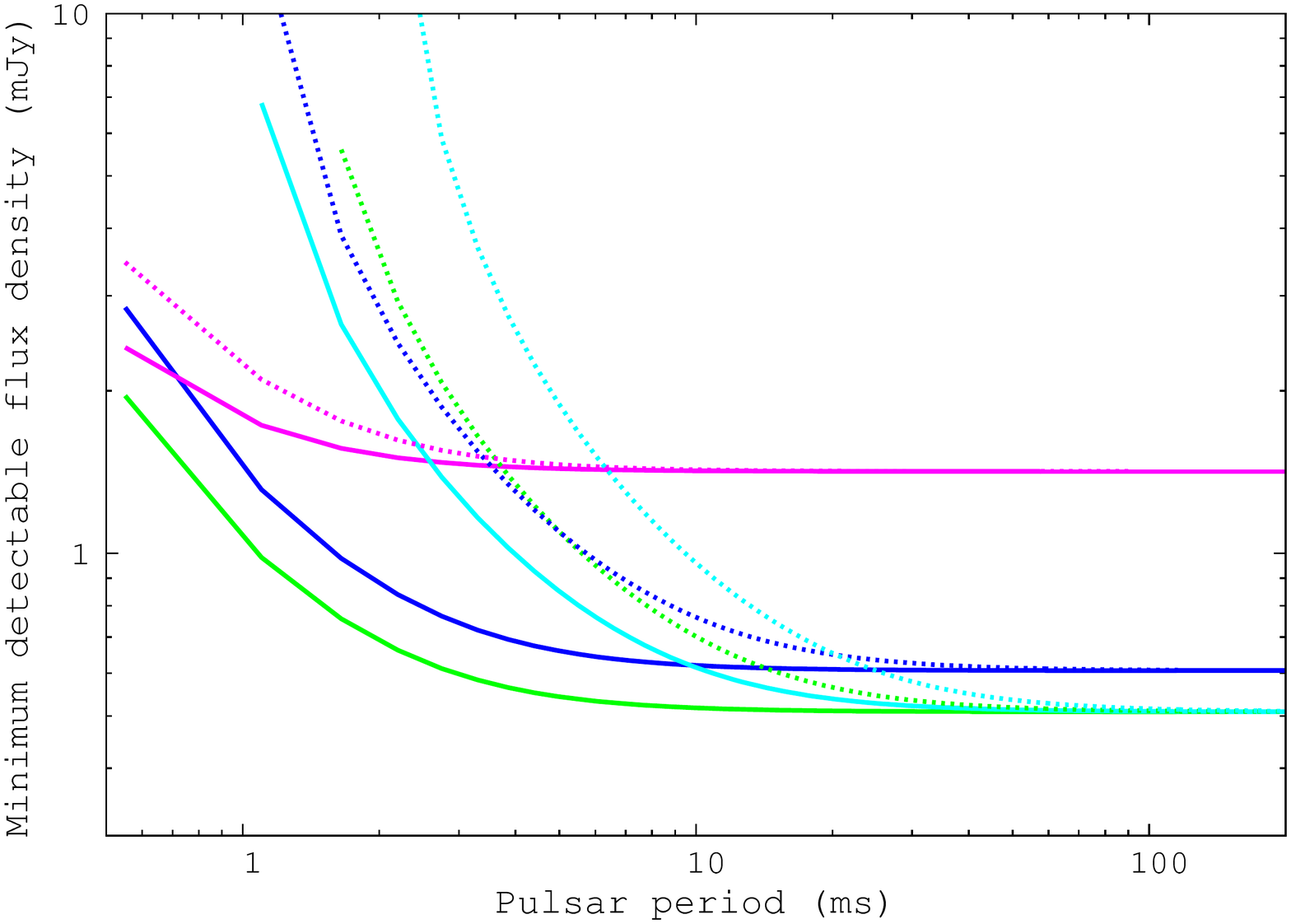}
    }
    \hfill
    \subfloat[\label{fig_sensitivity_msp}]{%
      \includegraphics[width=3.2in,angle=0]{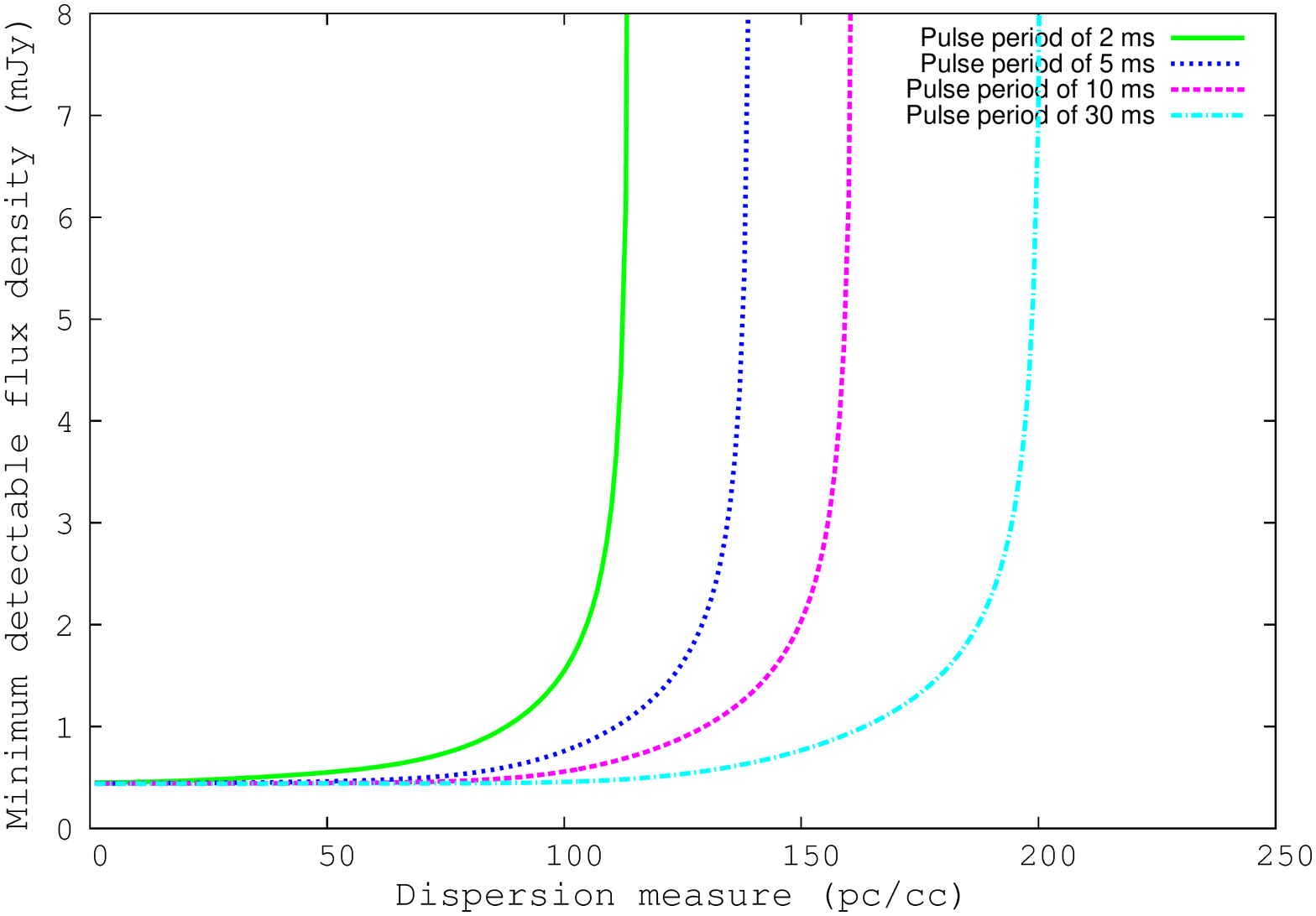}
    }
    \caption{(a) Estimated sensitivity of the GHRSS (green), GBNCC (blue), HTRU (magenta) and GMRT-FERMI (cyan) surveys. The solid lines are for DM 50 pc~cm$^{-3}$ and 
dashed lines for 100 pc~cm$^{-3}$. Sensitivities are scaled to 322 MHz with a spectral index of $-$1.7. (b) Estimated sensitivity of the GHRSS survey to MSPs as a 
function of DM for a range of pulse periods, 2 ms (green), 5 ms (blue), 10 ms (magenta) and 30 ms (cyan).}
  \end{figure}
\begin{figure}[h]
\centering
\begin{tabular}{c c}
\subfloat{\includegraphics[scale=0.4]{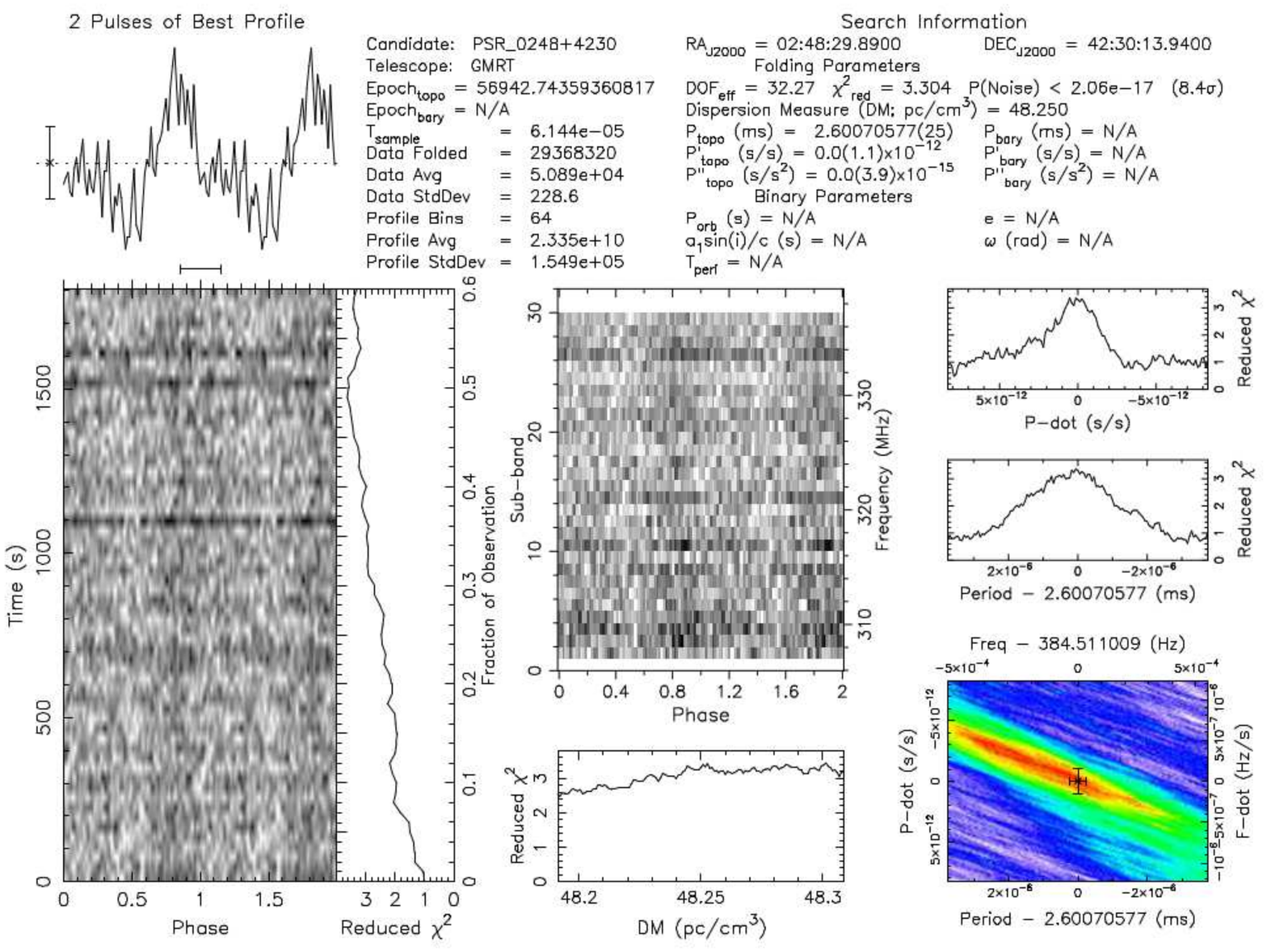}}
\\
\multicolumn{2}{c}
\subfloat{\includegraphics[scale=0.4]{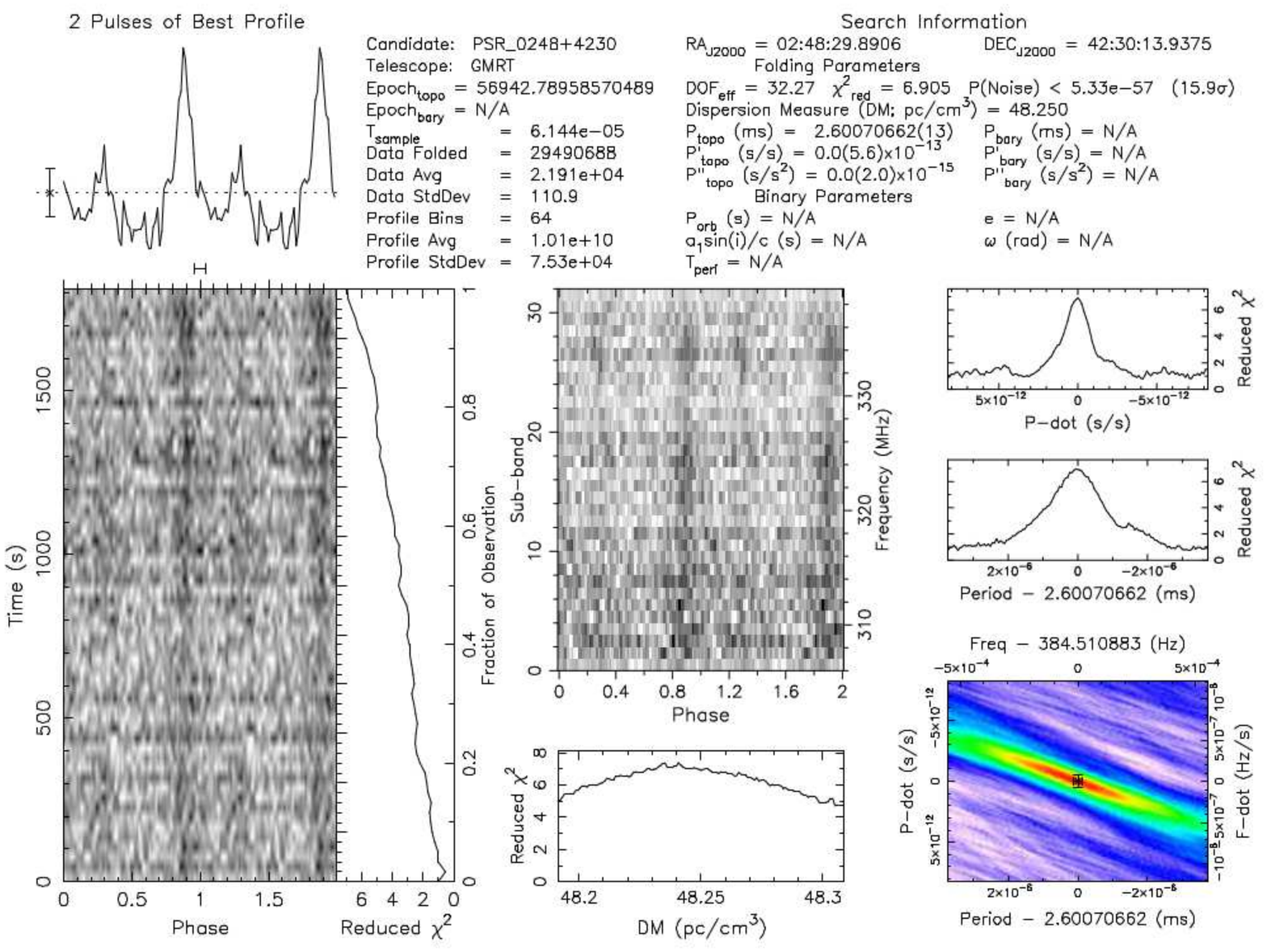}}
\end{tabular}
\caption{Output of {\sc presto} based GHRSS search pipeline for the MSP J0248$+$4230 to demonstrate the improvement in detection significance with the high resolution 
observing modes developed in the GHRSS survey. Top panel: with 32 MHz observing band split into 512 frequency channels recorded every 61.44 $\mu$s, Bottom panel: with 32 MHz 
observing band split into 2048 frequency channels recorded every 61.44 $\mu$s.}
\label{fig:512ch_2048_ch}
\end{figure}

A pulsar is detectable (with 5$\sigma$ detection significance) in a survey with the incoherent array if it exceeds some minimum flux density ($S_{\rm pulsar}$) 
that can be calculated using the radiometer equation:
\begin{equation}
S_{\rm pulsar} \sim 5 {{T_{\rm rec}+T_{\rm sky}}\over{G\sqrt{B N_{\rm p} N_{\rm a}t}}}
{\sqrt{{w}\over{P-w}}}
\end{equation}
where $T_{\rm rec}$ and $T_{\rm sky}$ are the temperatures of the receiver and sky respectively, $G$ the gain of individual antennas, 
$B$ the bandwidth, $N_{\rm p}$ the number of polarizations, $N_{\rm a}$ the number of antennas, $t$ the integration 
time, $w$ the effective pulse width (including all instrumental smearing), and $P$ the pulse period. 
As the telescope gain compensates for the larger bandwidths available at high frequencies ($\sim G_{low}/G_{high}>\sqrt{B_{high}/B_{low}}$),
it makes more sense to conduct surveys at lower frequencies as the larger beam-width greatly increases
the sky coverage ($\propto \nu^{-2}$) for a given dwell time ($t$) with similar sensitivity.

For the GHRSS survey we calculate a theoretical survey sensitivity of $\sim$ 0.5 mJy at 322 MHz (considering the radiometer equation)
for a 5$\sigma$ detection for 10\% duty cycle, with the GMRT incoherent array of gain 2.5 K/Jy ($\sim G\sqrt{N_{\rm p} {N_{\rm a}}}$) 
for 32 MHz bandwidth and considering a system temperature of 106 K. Figure \ref{fig_sensitivity} plots the estimated sensitivity of the GHRSS survey compared 
with GBNCC at GBT and HTRU at Parkes and the {\it Fermi}-directed survey at GMRT at DMs of 50 and 100 pc~cm$^{-3}$. 
Figure \ref{fig_sensitivity} illustrates that the GHRSS survey is the most sensitive survey in the GBNCC complementary sky.
Figure \ref{fig_sensitivity_msp} plots the estimated sensitivity of the GHRSS survey for MSPs as a function of DM. 
This indicates that for a 5 ms pulsar even at a DM of 100 pc~cm$^{-3}$ the estimated sensitivity is $\sim$ 0.75 mJy, which is only 1.5 times the predicted 
GHRSS sensitivity. Enhanced 
time and frequency resolution of GHRSS survey by a factor of 4 (Section \ref{sec:obs_sys}) enables an increased search sensitivity 
for MSPs up to a factor of 3 compared to the {\it Fermi}-directed MSP survey at the GMRT, which used 512$\times$0.0651 MHz channels 
sampled at 61.44 $\mu$s. This is validated by observing J0248$+$4230 in these observing modes (Figure \ref{fig:512ch_2048_ch}).
 With the newly implemented high resolution mode, the GHRSS survey is sensitive enough to detect MSPs at higher DMs as dispersion 
smearing is less than scatter broadening up to DM of 75 pc~cm$^{-3}$ (Figure \ref{fig_DM_scatter}). 
This will be useful considering 60\% of the survey region will have DM$<$ 100 pc~cm$^{-3}$ based on \cite{cordes02}. Thus the GHRSS survey 
modes are optimal for the targeted survey region. Figure \ref{fig_sensitivity} also illustrates that the GHRSS survey is the most sensitive survey in the 
GBNCC complementary sky. Table \ref{survey_comp} shows a comparison between the major off-Galactic plane surveys.

A transient event of 5 ms duration will be detected at 10$\sigma$ detection significance if it exceeds some minimum flux density ($S_{\rm transient}$)
calculated from the radiometer equation:
\begin{equation}
S_{\rm transient} \sim 10 {{T_{\rm rec}+T_{\rm sky}}\over{G\sqrt{B N_{\rm p} N_{\rm a}\tau}}}
\end{equation}
where $\tau$ is the duration of the burst, $T_{\rm rec}$ is 66 K and $T_{\rm sky}$ is 40 K. 
For the GHRSS survey we calculate a sensitivity of 1.6 Jy for a 10$\sigma$ detection limit for 5 ms transient millisecond bursts, 
considering weak scattering \citep{thornton13}.

\section{Re-detection of known pulsars}
\label{sec:redection}
\begin{figure}[htb]
\begin{center}
\includegraphics[width=7in,angle=0]{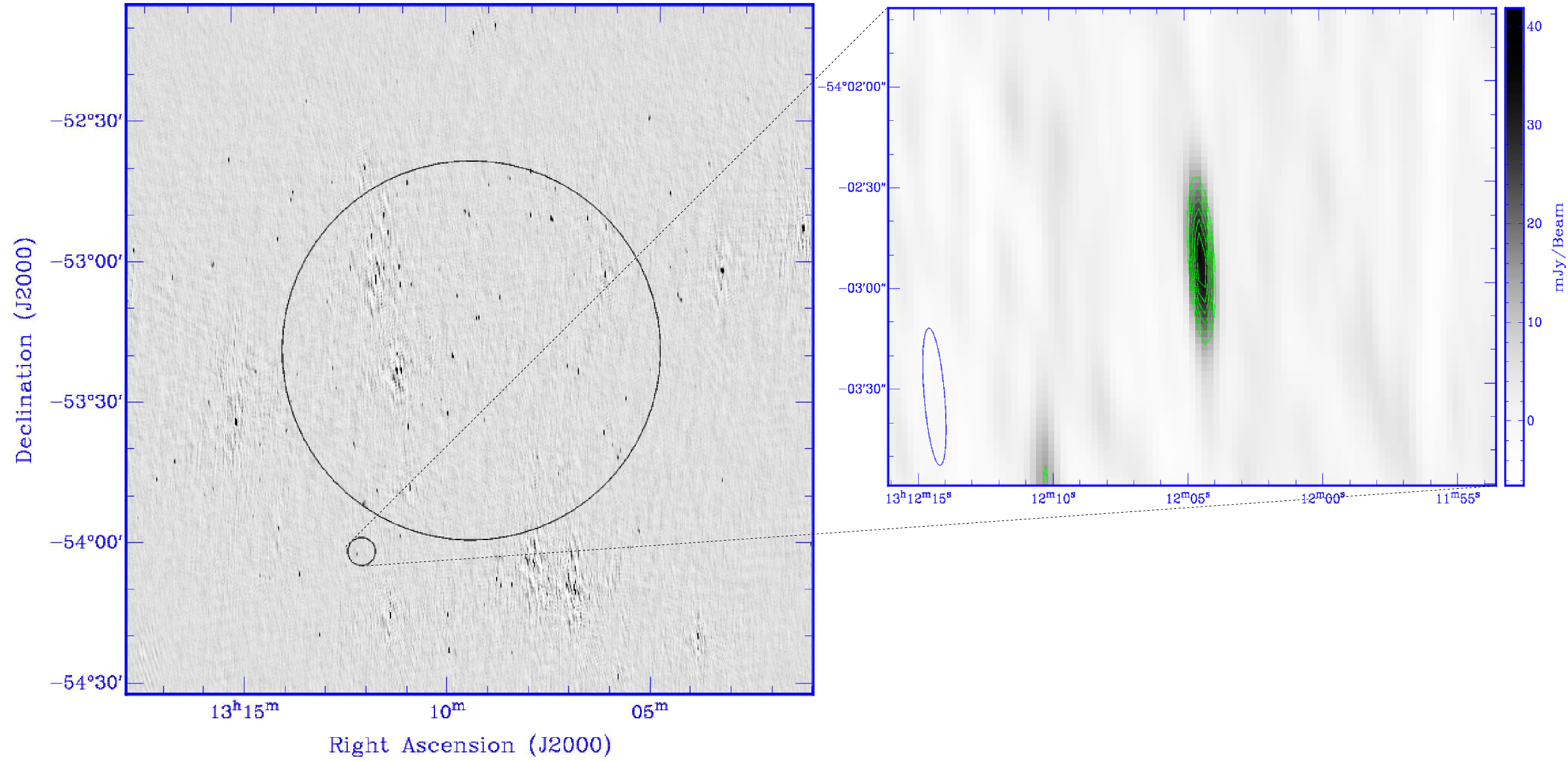}
\caption{Pulsar J1312$-$5402 was detected outside the HPBW (marked by bigger circle) and 49\arcmin~offset (marked by smaller circle) from the survey pointing centre with 17$\sigma$ 
pulsed detection significance. The interferometric flux density is 22 mJy. The right plot shows a small portion of the image containing the pulsar.}
\label{fig_redetection}
\end{center}
\end{figure}

\begin{figure}[htb]
\begin{center}
\includegraphics[width=4.5in,angle=0]{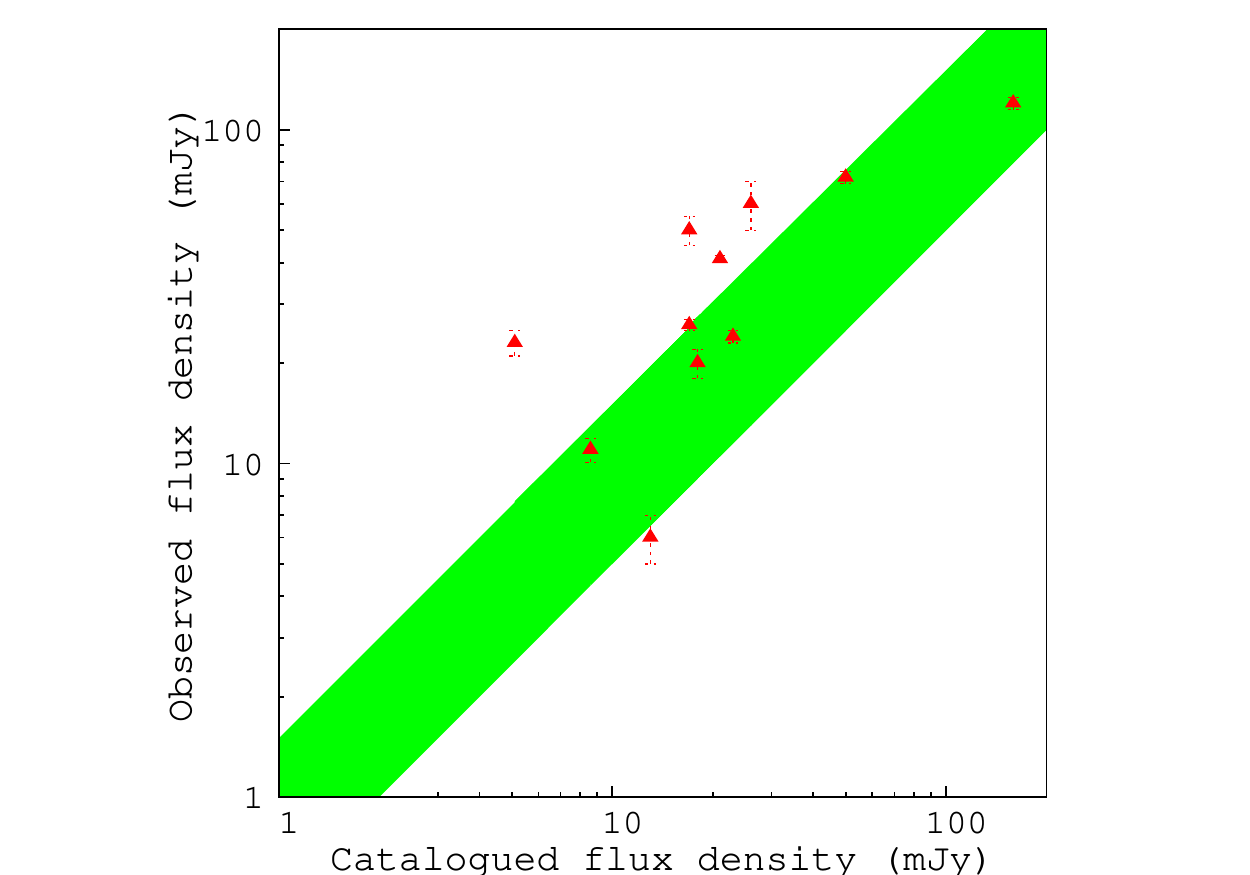}
\caption{Comparison of catalogued mean flux density at 322 MHz (extrapolated either by using the spectral index when available in the ATNF catalogue, or extracted 
considering 400 MHz mean flux density reported in the ATNF catalogue and a spectral index of $-$1.7) and observed interferometric flux density at 322 MHz for 12 re-detected pulsars.
The shaded part (green) represents a region with the observed flux density within $\pm$50\% of the catalogued flux densities.}
\label{atnf_sensitivity}
\end{center}
\end{figure}

Re-detection of known pulsars and comparison with the catalogued flux density helps us to evaluate the survey sensitivity. 
The pulsars detected within the measurable primary beam gain pattern (i.e. up to the first null at $\sim \pm$ 100\arcmin) of the GMRT 
antenna\footnote{http://www.ncra.tifr.res.in:8081/\~ngk/primarybeam/beam.html} were considered for flux density measurement. We had 
chosen only the highest signal-to-noise re-detection of a given pulsar, as some pulsars were re-detected in multiple adjacent scans. With 35\% of 
the GHRSS survey complete, we had 23 such re-detections of known pulsars. 
There are 30 known pulsars above 15$\sigma$ detection significance present within the observed GHRSS survey pointings. Most of 
the non detections of known pulsars correspond to the pulsars being at the edge of beam and/or presence of RFI.
We exploited the simultaneous availability of visibilities, to determine calibrated interferometric flux densities.
Imaging was done using a pipeline that was developed specifically for analysis of GMRT data (Kudale et al., in preparation). The raw interferometric
data were converted to FITS and then calibrated and flagged using the {\sc flagcal} pipeline \citep{prasad12,chengalur13}. 
We performed flux calibrations using flux densities of the observed phase calibrators as given in the VLA 
calibrator manual\footnote{https://science.nrao.edu/facilities/vla/docs/manuals/cal}.
The data were then imaged using a custom imaging pipeline built around utilities in the {\sc casa} package 
as well as the source identification package {\sc pybdsm} (paper in preparation). {\sc pybdsm} was used to identify regions with emission
in the dirty image which were set as the {\sc clean} boxes in the cleaning and self-calibration cycles. First a wide, low resolution image was
made (using mainly data from the central square antennas). We used this to identify sources that lie outside the half-power beam-width 
(HPBW) of the telescope. Then a high resolution image using all available antennas was made. The image covers the HPBW, but 
with extra ``flanking fields'' centred on all of the outlying sources identified in the low resolution image. The imaging and self calibration 
procedures typically included two rounds of phase-only self calibration followed by a final round of phase and amplitude calibration. Since most 
of the sources have been observed at very low elevation angles (as the fields in this survey were all southern fields) refractive effects can cause a significant
shift in the position of the sources. Wherever possible this shift was calibrated using catalogued observations at higher frequencies (e.g. using the
843 MHz SUMSS\footnote{http://heasarc.gsfc.nasa.gov/W3Browse/radio-catalog/sumss.html} catalogue in fields that overlap with that survey, 
or sources with known positions in the NED\footnote{https://ned.ipac.caltech.edu} database). After this correction we searched for 
emission around the known position of the pulsar. The measured flux density was then corrected for the primary beam attenuation to yield the final 
flux density of the pulsar. We could determine an interferometric flux density for 12 such re-detected fields having overlap with SUMSS/NED catalogued 
observations. Other re-detections could not be detected in the image due to increased noise near the pulsar position and/or significant flagging 
during imaging. Figure \ref{fig_redetection} shows the re-detection of PSR J1312$-$5402 at 49\arcmin~offset from the survey pointing centre, 
which clearly illustrates that pulsars of a few mJy can even be detected outside the HPBW of the survey pointing. Table \ref{redetection} 
lists the offset from the survey pointing centre, pulsed-SNR with primary beam correction and 
the corresponding imaging flux density of the re-detected pulsars. 
Similar to the GBNCC survey \citep{stovall14} we estimated the expected catalogued flux density at 322 MHz for each of these pulsars, either 
by using the spectral index when available in the ATNF catalogue, or by extrapolating 400 MHz mean flux density reported in the ATNF catalogue assuming 
a spectral index of $-$1.7. Large uncertainties could be associated with the catalogued 400 MHz flux densities as well with the assumption of $-$1.7 
spectral index, which indicates that the expected flux density estimates at 322 MHz have low accuracy. Figure \ref{atnf_sensitivity} compares 
the observed interferometric flux density with the extrapolated catalogued flux density. The majority of the detections have observed flux densities within $\pm$50\% 
of the expected values (marked by green shaded region). In conjunction to the uncertainties involved in estimating the expected catalogued flux density, 
the observed sensitivity degradation for some of the re-detections could also be caused due to scintillation or significant flagging due to RFI. 

\section{Discoveries}
\label{sec:discoveries}
\begin{figure}
\includegraphics[width=5.5in,angle=0]{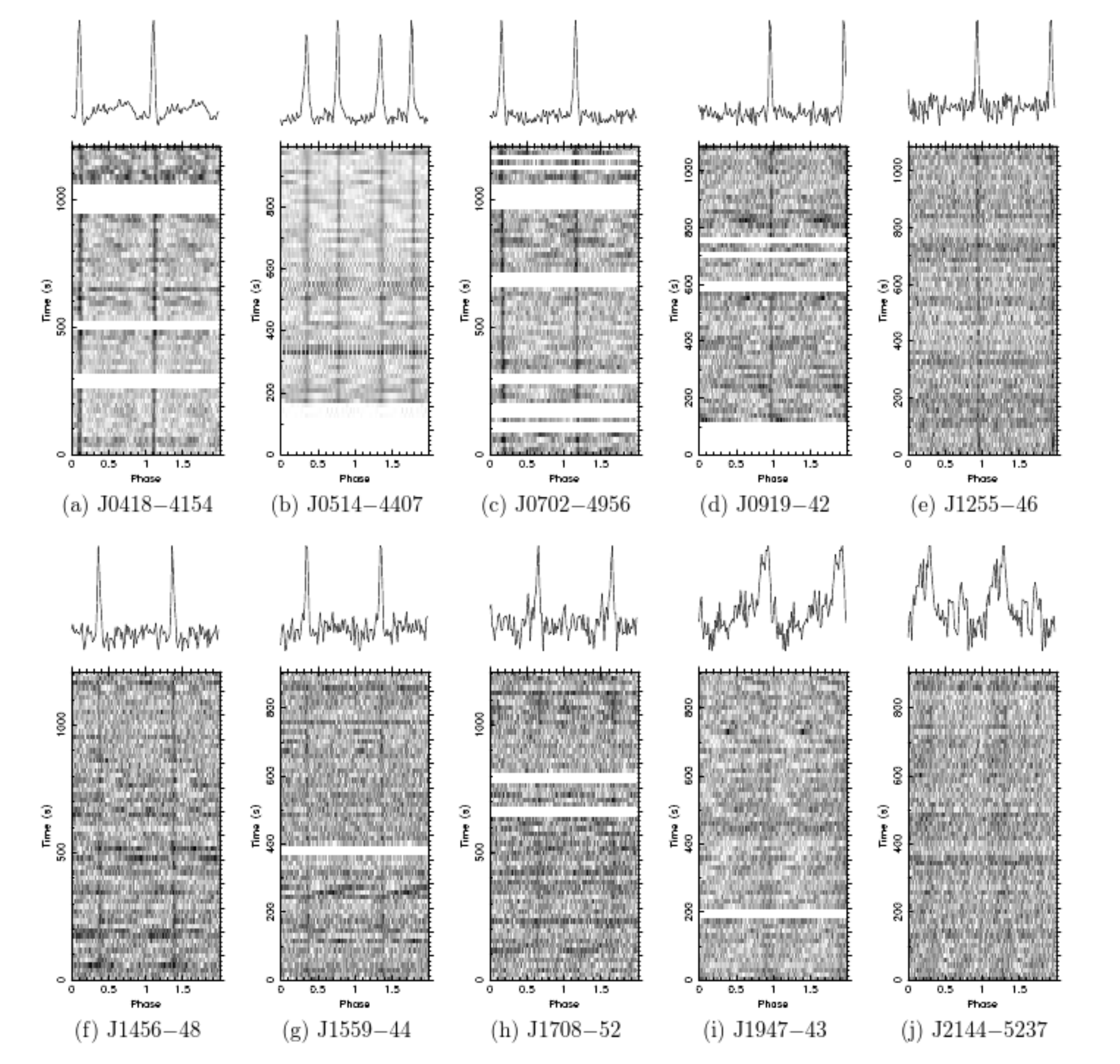}
\caption{Discovery plots of the ten pulsars discovered in the GHRSS survey. The observations were performed in the incoherent
array mode. The folded profiles are plotted (twice) in the top panel of each plot. Signal strength as a function of
rotational phase is plotted in the bottom panel.
Severe RFI occasions are masked, however as can be seen in some cases significant RFI remains.}
\label{fig:discoveries}
\end{figure}

In this paper we report pulsar and transient search results from 35\% completion of the GHRSS survey covering 1000 deg$^2$ of sky.
With the periodicity search we discovered 10 pulsars including one millisecond pulsar in a binary system.
Figure \ref{fig:discoveries} shows the discovery profiles and 
Table \ref{discovery} lists the periods, dispersion measures and discovery flux density values. Though some of the 
newly discovered pulsars are reasonably strong and might have been missed by previous surveys (e.g. J0418$-$4154, J0702$-$4956 and J0514$-$4407), we 
also discovered a few weak pulsars, indicating the highly sensitive nature of the survey even 
in the presence of some residual RFI after flagging. At the discovery epoch the newly discovered MSP had an estimated 
flux density of 1.6 mJy which is close to the theoretical sensitivity considering the system temperature of 109 K \citep{haslam82} 
and 30\% duty cycle.  We confirmed all the discoveries in subsequent epochs of observations. The GHRSS discoveries have 
dispersion measures ranging from 15.4 pc~cm$^{-3}$ (for PSR J0514$-$4407) up to 133 pc~cm$^{-3}$ 
(for PSR J1456$-$48) and periods from 5.04 ms (for PSR J2144$-$5237) up to 1169.89 ms (for PSR J1559$-$44).   
Most of the newly discovered pulsars (e.g. J0418$-$4154, J0702$-$4956, J0919$-$42, J1255$-$46, J1456$-$48, J1559$-$44 and J1708$-$52) have single 
peaked profiles. PSR J1947$-$43 has a relatively wide profile and there is a hint of a double peaked profile. For PSR J0514$-$4407 a double 
component pulse profile is observed. The profile shapes were confirmed with multi-epoch observations. The 52.04 ms rotational 
period of PSR J1255$-$46 puts it in the domain of young pulsars or double neutron star relativistic binaries. A more detailed 
study of this pulsar is in preparation. The 5.04 ms MSP J2144$-$5237 has a multi-component profile. The right panel of Figure \ref{fig_incoh_coh} shows the 
sensitive coherent array observations aided by precise localisation (Section \ref{sec:localisation}). The flux density of 
this MSP varies dramatically due to scintillation, indicating the need for re-observations of candidate MSPs with 
longer integrations. 
A drift over Fourier frequency by 4$\times$10$^{-4}$ Hz, which corresponds to a period derivative 
of $-$1.4$\times$ 10$^{-12}$ s s$^{-1}$ or a line-of-sight acceleration of 1.4 m s$^{-2}$ observed 
for J2144$-$5237 over two hours, indicates that this MSP is in binary system. 
PSR J2144$-$5237 could be a good timer for inclusion in the International Pulsar Timing Array. 

The area normalised discovery rate of the GHRSS survey is 0.01 pulsars per deg$^2$ which is higher than the 0.006 pulsars per deg$^2$ 
achieved with the GBNCC survey (Section \ref{sec:survey_prediction}). It may be that this higher discovery rate is purely due to low number 
statistics at this stage or could perhaps represent an underestimate of the sensitivity of the GMRT in the incoherent mode or 
perhaps an overestimate of the GBNCC sensitivity. It may be useful to survey an overlapping region of sky to enable 
a more direct comparison between these low frequency surveys. However the GHRSS pulsar discovery rate is lower than 
the prediction by {\sc psrpoppy} prediction (Section \ref{sec:survey_prediction}) like many of the other surveys.

We have not detected any FRB or RRAT with the single pulse processing in the GHRSS survey. However, considering the 
FRB rate predicted in Section \ref{sec:survey_prediction} we should have detected only about 2$^{+1}_{-1}$ FRBs in this survey 
region based on a flat spectral index consideration. Absence of which is not constraining enough considering factor of few 
lower FRB prediction by recent studies \citep{rane15, champion15, petroff14}.

\begin{table*}
\begin{center}
\caption{Parameters of the pulsars discovered in GHRSS survey}
\vspace{0.3cm}
\label{discovery}
\begin{tabular}{|l|c|c|c|c|c|c|c|c|c|c|c|c|c|c|c|c}
\hline
Pulsar name    & Period  & Dispersion measure & Detection significance & Flux density$^\dagger$ \\
               & (ms)    & (pc~cm$^{-3}$)            & ($\sigma$)             & (mJy)  \\\hline
PSR J0418$-$4154 & 757.11  & 24.5             & 50                     & 10.3                        \\
PSR J0514$-$4407 & 302.2   & 15.4             & 42                     & 9.7  \\
PSR J0702$-$4956 & 666.66  & 98.7             & 30                     & 15.7     \\
PSR J0919$-$42 & 812.6              & 57.8                  & 19       & 6.4     \\
PSR J1255$-$46 & 52.0               & 42.9               & 12          & 0.8 \\
PSR J1456$-$48 & 536.81  & 133.0              & 15                     & 1.2   \\
PSR J1559$-$44 & 1169.89 & 122.0              & 8                      & 1.7    \\
PSR J1708$-$52 & 449.62  & 102.6              & 9                      & 1.4  \\
PSR J1947$-$43 & 180.94  & 29.9               & 17                     & 4.7   \\
PSR J2144$-$5237 & 5.04    & 19.0               & 9                    & 1.6 \\\hline
\end{tabular}
\end{center}
$^\dagger$ : Flux density is without primary beam correction.
\end{table*}

\begin{table*}
\begin{center}
\caption{List of 23 re-detected pulsars in the GHRSS survey. Offsets from the GHRSS beam centre, primary beam corrected pulsed SNR and interferometric flux densities (for 12 pulsars, the others either did not had overlapping SUMSS/NED observations, or could not be detected in the image due increased RMS near the pulsar position and/or significant flagging during imaging) are tabulated.}
\vspace{0.3cm}
\label{redetection}
\begin{tabular}{|l|c|c|c|c|c|c|c|c|c|c|c|c|c|c|c|c}
\hline
Pulsar name     & offset (\arcmin)   &   Pulsed-SNR   & Interferometric Flux density (mJy) \\\hline
J0255$-$5304    & 25               &   31            &   - \\
J0745$-$5353    & 72               &   272            &   -  \\
J0809$-$4753    & 69               &   176    &   - \\
J0842$-$4851    & 53               &   60   &   - \\
J0840$-$5332    & 53               &   30    &   - \\
J0843$-$5022    & 0.0              &   11    &   12$\pm$1 \\
J0905$-$5127    &  9               &   28    &   - \\
J0907$-$5157    & 51               &   130    &   72$\pm$3  \\
J0924$-$5302    & 26               &   77    &   26$\pm$1 \\
J0934$-$5249    & 54               &   46    &   60$\pm$10 \\
J0955$-$5304    & 62               &   132    &   - \\
J1003$-$4747    & 2.5              &   82    &   11.0$\pm$0.9  \\
J1036$-$4926    & 43               &   52    &   6.0$\pm$2 \\
J1045$-$4509    & 33               &   50    &   24$\pm$1  \\
J1143$-$5158    & 40               &   64    &    - \\
J1240$-$4124    & 36               &   93    &   23$\pm$2  \\
J1312$-$5402    & 58               &   43    &   41$\pm$1  \\
J1320$-$5359    & 79               &   428    &  20$\pm$2\\  
J1355$-$5153    & 47               &   114   &   50$\pm$5  \\
J1544$-$5308    & 15               &   36   &   -  \\
J1559$-$4438    & 3.7              &   59    &   120$\pm$5 \\
J1902$-$5105    & 75               &   159    &   - \\
J2241$-$5236    & 13               &   71    &   - \\\hline
\end{tabular}
\end{center}
\end{table*}

\section{Localisation of the GHRSS discoveries}
\label{sec:localisation}
We utilised the interferometric imaging capability of the GMRT to localise pulsars and transient events. Such 
localisation in the image plane is very important for FRBs for host galaxy identification and multiwavelength follow up. \\ 
\subsection{Localisation of the pulsars with gated correlator}
A gating interferometer is an excellent tool to improve the noise statistics in order to increase the 
detection significance towards a pulsed signal. The development of a coherently dedispersed 
gated correlator for the GMRT is detailed in \cite{roy13}. In this design, the visibility 
time-series were derived off-line from the recorded raw voltage data and were binned using multiple 
gates and on each gate they were folded using the best-fit topocentric rotational model, derived from 
the synchronously generated incoherent beam data. We improved the noise statistics further by subtracting 
the visibilities of adjacent gates. The ON-OFF subtracted visibilities were then calibrated and flagged 
using {\sc flagcal} pipeline \citep{prasad12,chengalur13}. The ON-OFF image was made using the {\sc aips} 
package, where the pulsar is unambiguously localised as a point source. For pulsars with 10\% duty cycle, 
we expect to get a factor of 3 improvement in detection significance in the gated image plane compared to the 
continuum plane. The positional uncertainty therefore reduces from 80\arcmin~(half-power 
beam-width at 322 MHz) to $<$20\arcsec~(typical synthesized beam used in the image made at 322 MHz).

\begin{figure}
\begin{center}
\includegraphics[width=5in,angle=0]{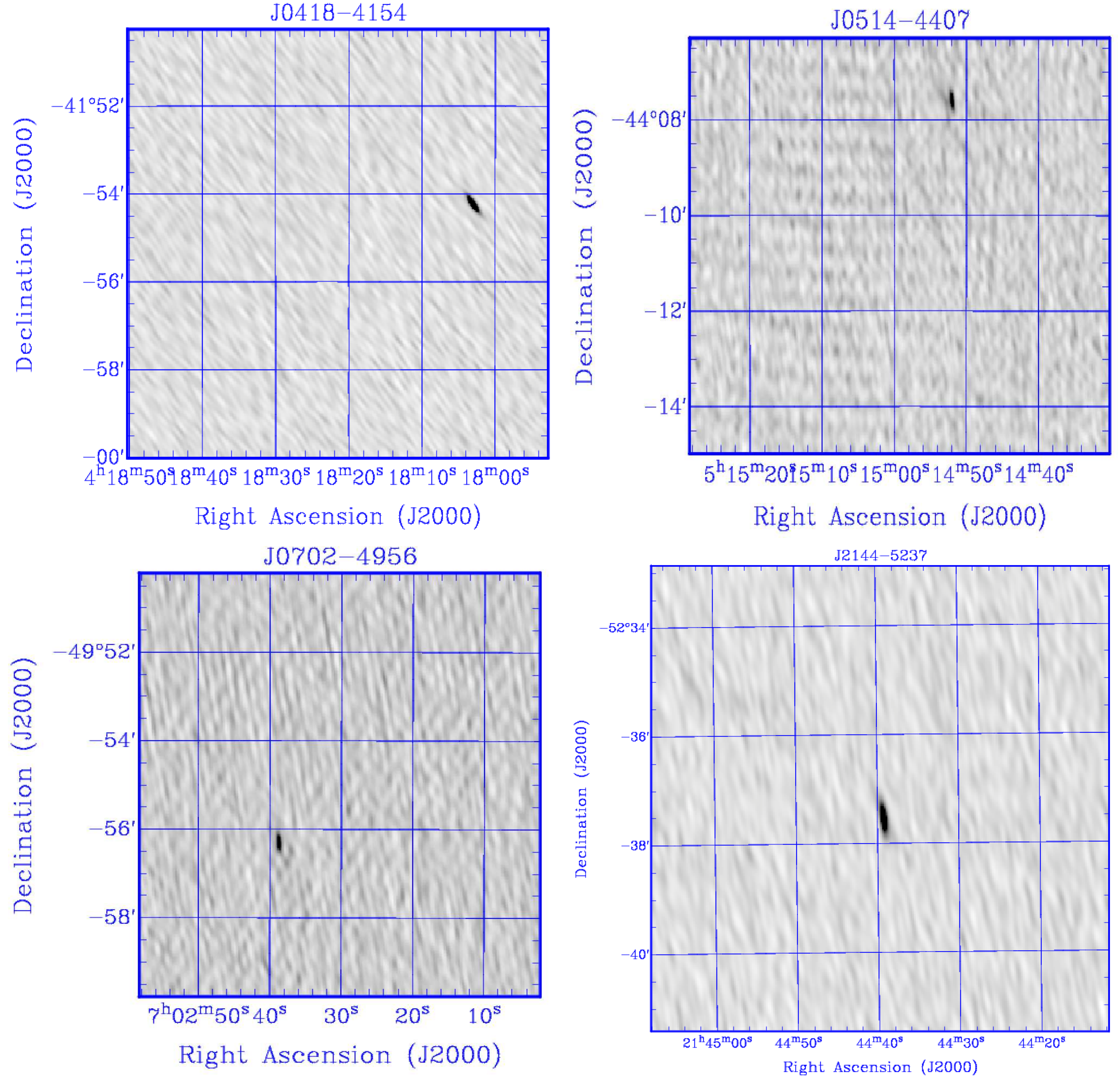}
\caption{Localisation images for 4 pulsars discovered in the GHRSS survey: J0418$-$4154, J0514$-$4407, J0702$-$4956 and J2144$-$5237. The offsets of the pulsar from the pointing centre is given in Table \ref{gating_param}. Since J0514$-$4407 and J2144$-$5237 are at large offsets from the pointing centre, we show a smaller field closer to the pulsar.}
\label{fig_localisation}
\end{center}
\end{figure}

We recently started follow up of these pulsars using the coherently dedispersed gated correlator which 
resulted in the successful localisation of 4 pulsars, namely: J0418$-$4154, J0702$-$4956, J0514$-$4407 and J2144$-$5237 with an 
accuracy of $<$10\arcsec~(positions are listed in Table \ref{gating_param}). Figure \ref{fig_localisation}
plots the ON-OFF gated images of these 4 pulsars. Such timing independent, rapidly determined and fairly precise positions allow us
to use sensitive coherent beams (up to 5$\times$ the sensitivity of the incoherent beam) of the GMRT for follow up timing observations, 
substantially reducing the use of telescope time. Further improvement of the detection significance is seen while using the 
coherent array, as it is better resistant to RFI. Right panel of Figure \ref{fig_incoh_coh} shows the huge improvement in sensitivity achieved with 
coherent array observations for PSR J2144$-$5237. Moreover, such a priori astrometric models allow, rapid convergence in pulsar 
timing as discussed in Section \ref{sec:timing}. 
 
\subsection{Localisation of FRBs in the image plane}
\label{FRB_loc}
In addition to localisation of the pulsars, simultaneous availability of the visibility data during the high resolution 
survey provides us the opportunity to localise the transient events detected in time-domain searches.
All 17 known FRBs have been discovered with single dish telescopes and interferometric detection of an FRB is of significant 
importance. This technique is successfully utilised in the study of the eclipsing pulsars with the GMRT (e.g. PSR J1227$-$4853 \citep{roy15}).
An example of interferometric localisation of a re-detection of a known pulsar in the GHRSS survey (Figure \ref{fig_redetection}) also 
illustrates the pipeline that will be followed for localisation of possible transient detection in the GHRSS survey. 
Assuming that interferometric gain is similar to coherent array gain (i.e. 5 times of incoherent array), FRBs with flux density $>$~3 Jy could be localised
at $>$ 5$\sigma$ significance in the image plane from simultaneously recorded visibilities at 2 s resolution.

\begin{table}
\begin{center}
\caption{Parameters obtained from the gated imaging}
\vspace{0.3cm}
\label{gating_param}
\begin{tabular}{|l|c|c|c|c|c|c|c|c|c|c|c|c|c|c|c|c}
\hline
PSR         &Gated J2000      &Offset from     &Number     &Observing      &Gated      &Gated \\
            &position         &pointing        &of gates   &duration       &flux density   &SNR   \\
            &(\arcsec)&centre        &           & (s)          &(mJy)      &     \\\hline
J0418$-$4154  &04$^\mathrm{h}$18$^\mathrm{m}$02\fs88(1\farcs4);&7.4\arcmin    & 24     &900 &88        &9   \\
            &$-$41\degr54\arcmin11\farcs89(7\farcs8)             &                &         &   &           &    \\
J0514$-$4407&05$^\mathrm{h}$14$^\mathrm{m}$51\fs84(1\farcs04);&32\arcmin      & 21     &900  &20        &5    \\
            &$-$44\degr07\arcmin06\farcs51(8\farcs4)              &               &         &   &           &  \\
J0702$-$4956&07$^\mathrm{h}$02$^\mathrm{m}$38\fs54(1\farcs2);&6.2\arcmin        & 16      &900  &29       &6  \\
            &$-$49\degr56\arcmin56\farcs27(7\farcs8)           &                  &         &   &           &    \\
J2144$-$5237&21$^\mathrm{h}$44$^\mathrm{m}$39\fs2(0\farcs7);&28\arcmin        & 10      &3600  &6       &12  \\
            &$-$52\degr37\arcmin32\farcs17(3\farcs8)        &                  &         &   &           &    \\\hline
\end{tabular}
\end{center}
 \end{table}

\section{Timing and follow up of the GHRSS discoveries}
\label{sec:timing}
\begin{figure}[!ht]
    \subfloat[\label{fig:512ch}]{%
      \includegraphics[width=3.5in,angle=0]{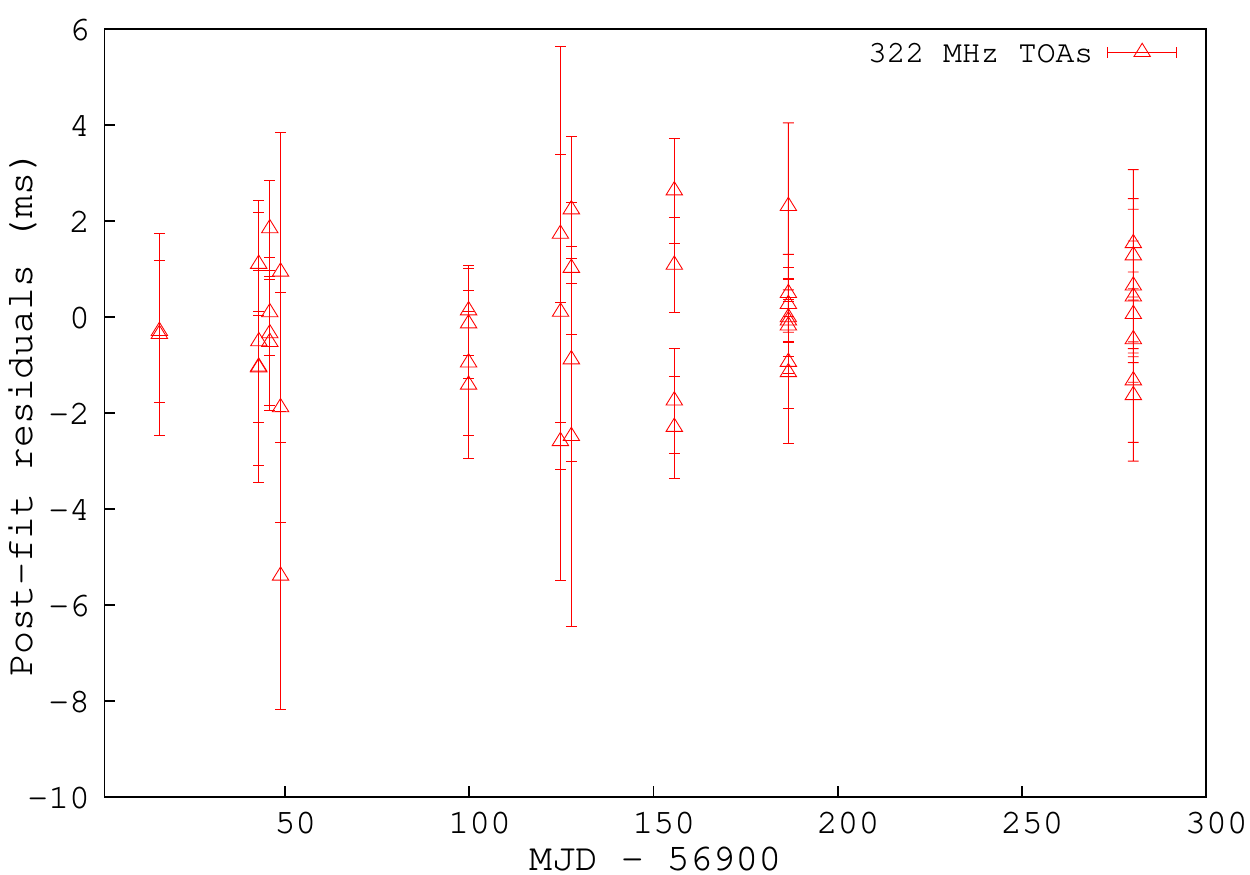}
    }
    \hfill
    \subfloat[\label{fig:2048ch}]{%
      \includegraphics[width=3.5in,angle=0]{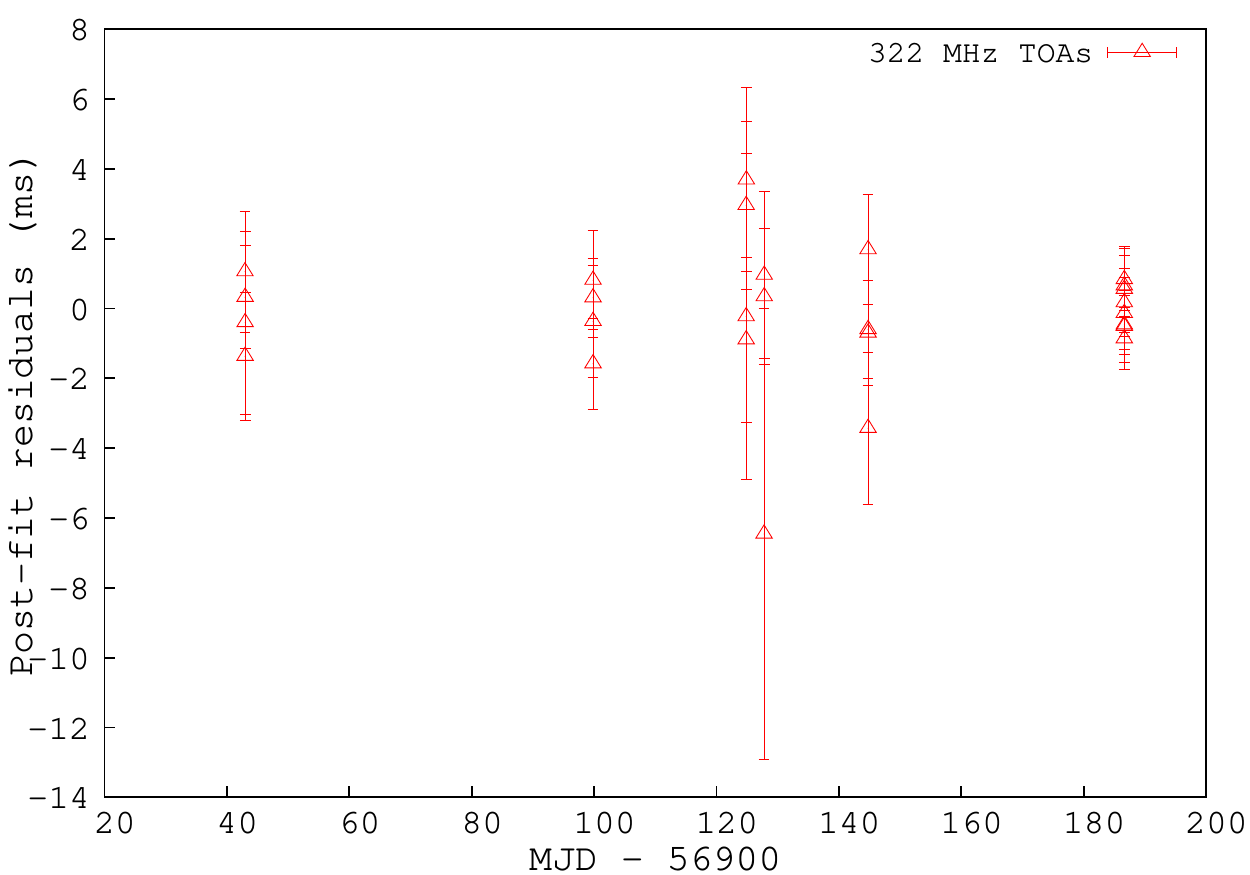}
    }
    \caption{Post-fit timing residuals for J0418$-$4154 (left panel) and J0702$-$4956 (right panel). Multiple TOAs are derived for each observing epoch. Post-fit residual is 1.1 ms for J0418$-$4154 and 1.0 ms for J0702$-$4956.}
    \label{fig_residuals}
  \end{figure}

\begin{figure}
\begin{center}
\includegraphics[width=0.7\textwidth,angle=0]{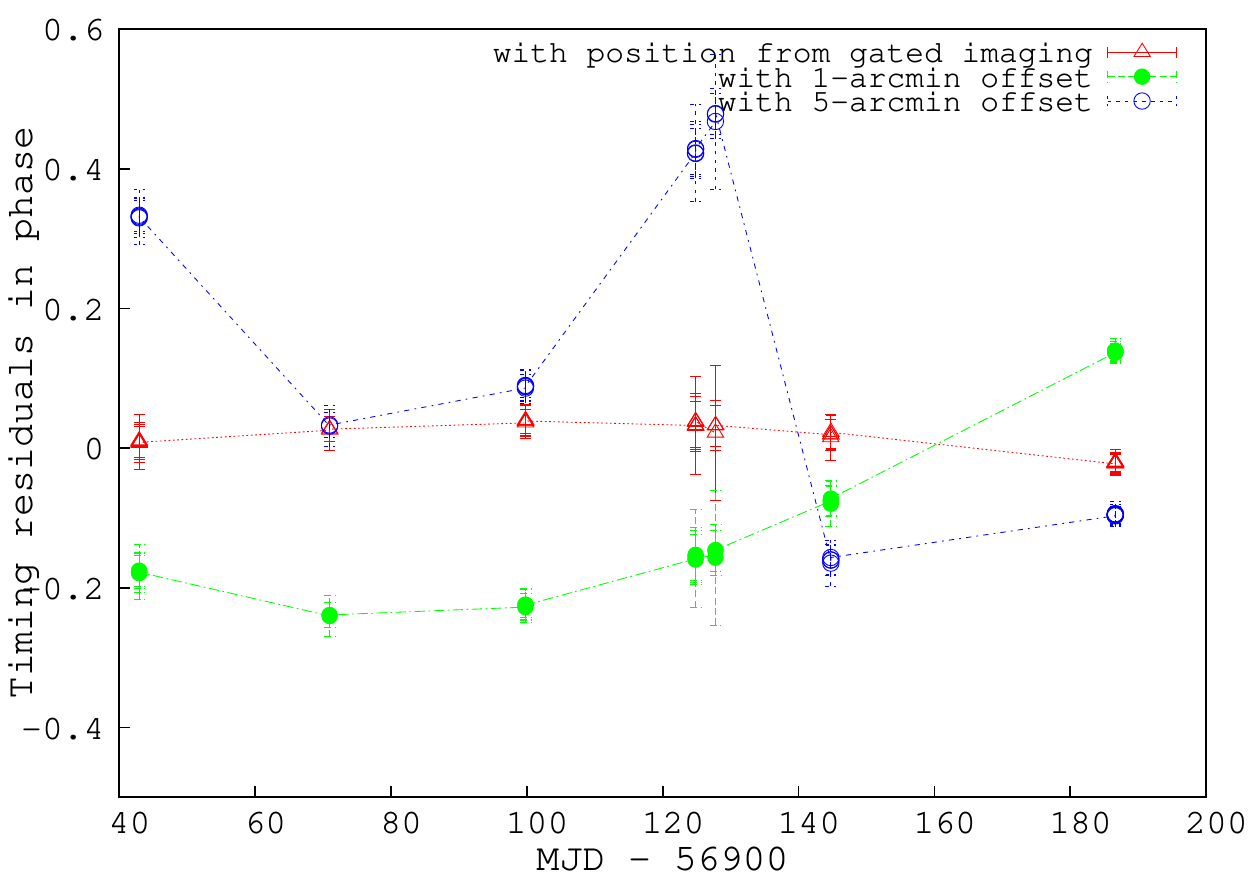}
\caption{The post-fit timing residual for PSR J0702$-$4956 fitted with three timing models having different astrometric positions,
the gated interferometric position (solid red line), 1\arcmin~ offset (green dashed line) and 5\arcmin~ offset (blue dotted line)
from the gated interferometric position. The errors on TOAs are magnified by 10$\times$ for better visualisation. A $\pm$0.5 phase 
wrap is noticed for the timing residual with 5\arcmin~ position offset.}
\label{timing_J0702-4956}
\end{center}
\end{figure}

We are continuing regular timing follow up of the GHRSS pulsars, with the coherent array of the GMRT, to determine 
their rotational and binary parameters (for the MSP) and place them on the $P$-$\dot{P}$ diagram. 
Long term timing will also allow us to probe the possible optical and $\gamma$-ray counterparts of the newly 
discovered pulsars. The timing-independent rapid astrometric measurement of newly discovered pulsars using gated imaging helps to
break the degeneracy in the timing fit caused by co-variance between pulsar position and rotational parameters.
We successfully demonstrated this by timing two of the localised pulsars, J0418$-$4154 
and J0702$-$4956, using sparsely spaced TOAs over a span of $\sim$ 200 days. Considering reasonable fractional 
bandwidth of 10\% at 322 MHz, we determined the DM using timing fits of sub-band TOAs. The timing ephemeris of both the pulsars are 
given in Table \ref{tab:residuals}. The post-fit residuals are shown in Figure \ref{fig_residuals}. 
Residuals for J0702$-$4956 are fitted with three timing models having different astrometric positions: the gated 
interferometric position, 1\arcmin~offset and 5\arcmin~offset from the gated interferometric position, which is illustrated in Figure \ref{timing_J0702-4956}.
The residuals increase as the position offset increases and the residuals are wiggled in phase at 5\arcmin~offset. This warrants more closely 
spaced and regularly sampled TOAs to overcome the phase ambiguity caused by position error. Thus a priori knowledge of a pulsar 
position from the gated-image is extremely important, considering that the discovery position can be anywhere within 
the HPBW ($\pm$ 40\arcmin) or even outside the HPBW (e.g. Figure \ref{fig_redetection}).   
  
\begin{table}
\centering
\caption{Timing parameters of PSR J0418$-$4154 and J0702$-$4956
\label{tab:residuals}}
\begin{tabular}{|l|c||c|c|}
\hline
Parameters & J0418$-$4154 & J0702$-$4956 \\
  \hline
  \multicolumn{3}{c}{Gated imaging position} \\
  \hline
Right ascension(J2000)\dotfill & 04$^\mathrm{h}$18$^\mathrm{m}$02\fs88$\pm$1\fs4 & 07$^\mathrm{h}$02$^\mathrm{m}$38\fs54$\pm$1\fs2 \\
Declination(J2000)\dotfill     & -41\degr54\arcmin11\farcs89$\pm$7\farcs8          & -49\degr56\arcmin56\farcs27$\pm$7\farcs8 \\
  \hline
  \multicolumn{3}{c}{Parameters from radio-timing$^a$} \\
  \hline
Right ascension(J2000)\dotfill & 04$^\mathrm{h}$18$^\mathrm{m}$04\fs2(2)         &  07$^\mathrm{h}$02$^\mathrm{m}$39\fs3(2)\\
Declination(J2000)\dotfill     & $-$41\degr54\arcmin10\farcs8(6)                 & $-$49\degr56\arcmin34\farcs2(6)\\
Pulsar frequency $f$(Hz)\dotfill  & 1.3207965590(1)                               &  1.501515953(1)\\
Pulsar frequency derivative $\dot{f}$ (Hz s$^{-1}$)\dotfill & $-$1.7 (1)$\times$10$^{-15}$ &  0.00\\ 
Period epoch(MAD)\dotfill         & 57055                                    &  57086\\
Dispersion measure $\mbox{DM}$(pc~cm$^{-3}$)\dotfill & 24.54(8)                    &  98.7(1) \\
Span of timing data(MAD)\dotfill  & 56915.88--57180.25                            &  56942.95--57086.64 \\
Number of TOAs\dotfill             & 44                                                       &  29 \\
Post-fit residual rms(ms)\dotfill & 1.1                                         &  1.00 \\
Reduced chi-square\dotfill & 1.05                                                  &  0.7 \\\hline
\end{tabular}
\tablenotetext{a}{Errors in the last digit are in parentheses}
\end{table}

Following the localisation of a GHRSS pulsar we routinely check for nearby sources from existing optical\footnote{http://simbad.u$-$strasbg.fr/simbad/sim$-$fcoo} 
and high-energy catalogues\footnote{http://fermi.gsfc.nasa.gov/ssc/data/access/lat/4yr\_catalog/}. For one of the GHRSS discovery pulsar, J0514$-$4407, we found 
a possible counterpart 3FGL J0514.6$-$4406, with a position uncertainty of about 0.16\degr. This is only 1.8\arcmin~from the pulsar, so well within predicted 
$\gamma$-ray position. However, none of the other pulsars listed in Table \ref{gating_param} have 2FGL/3FGL counterparts. We plan to fold the {\it Fermi} photons 
with the timing model of the newly discovered pulsars to check if any of these pulsars exhibit $\gamma$-ray pulsations. This will provide input towards the radio-loud versus radio-quiet 
$\gamma$-ray pulsar ratio, which is a reasonable differentiator between $\gamma$-ray population models. 

\section{Relevance to the SKA}
\label{sec:SKA}
Discoveries of pulsars, in particular MSPs, are hindered by their radio faintness that demands deeper searches with larger 
telescopes and we are reaching the limit of what is possible for fully steerable single dishes. Large arrays of many smaller 
telescopes are the future for large radio telescopes leading ultimately to the world's largest telescope the SKA. 
The SKA's wide field-of-view, high sensitivity, multi-beaming and sub-arraying capabilities, coupled with advanced 
pulsar search backends, will result in the discovery of a large population of pulsars \citep{keane15}. 
As it is currently the largest telescope in the metre-wavelength regime, the GMRT, an SKA pathfinder telescope, 
is the prototype for the SKA in many ways, and provides an excellent test bed for new techniques.
Tools for efficient RFI mitigation, optimised search techniques, synchronous beam-formation and imaging will provide 
vital input to the SKA. In particular the high resolution GHRSS data are being used for testing the optimised 
dedispersion and de-acceleration plan of the pulsar search sub-element (PSS) of SKA. The periodicity search for the GHRSS 
needs an 8 million point transform, similar to the requirement of the SKA. Moreover, the rapid convergence in pulsar timing model possible with 
precise localisation of the pulsars with gated imaging described in Section \ref{sec:localisation} and \ref{sec:timing}, demonstrates 
a feature that may be applicable in the SKA.    

The GHRSS survey is at a frequency that overlaps with the SKA1 low frequency instrument (SKA1$-$Low) and considering 
SKA1-Low will have sensitivity $\sim$ 0.05 mJy at 350 MHz \citep{keane15} (i.e. 10 times GHRSS sensitivity) 
all the pulsars discovered with the GHRSS survey will be able to be timed and studied with SKA1$-$Low, providing an
excellent opportunity to prepare for searches and timing for this next generation telescope. The GPU based processing
pipeline implemented by us, {\sc biforst}, performs a time-domain acceleration search on pulsar survey data, and the
processing steps roughly follow those of the PSS time-domain processing pipeline. The GHRSS survey data are concurrently 
 being processed by time-domain acceleration search using GPU and frequency-domain acceleration search using CPU, and 
thus provides detailed comparison between the algorithms, which can be a vital input to SKA choices to decide the 
appropriate search methodology. These results will provide very valuable input into directing the prototyping and 
design effort for pre-construction stage. In future for the GHRSS survey we plan to use the baseband recording facility of 
the GMRT Software Backend followed by multi-pixel coherent beamformation (Roy et al. 2012). 
Pixelisation of the full field-of-view while scanning the extreme southern sky with the GMRT will provide 3$\times$ sensitivity improvement 
and 20$\times$ increase in compute cost. For the extreme southern sky we will require 300 beams which is $\sim$ 10\% compute cost of SKA pulsar search. 

\section{Conclusion}
\label{sec:conclusion}
In this paper we described the system configuration and initial discoveries from the GHRSS survey. We described the high 
resolution observing modes developed for this survey and the data processing pipeline for periodicity and single pulse search. 
Our survey sensitivity is comparable to other ongoing low frequency surveys, and we are targeting 
a portion of sky which has not been surveyed at lower frequencies like 322 MHz for the last two decades. With 35\% completion of the GHRSS survey covering 1000 
deg$^2$ of sky we discovered 10 pulsars including 1 MSP which corresponds to one of the best pulsar per deg$^2$ 
 discovery rates of any survey off the Galactic plane. We also re-detected 23 previously known pulsars with the expected detection significance, 
which were in-beam pulsars, not necessarily at the pointing centre. The simultaneous time-domain search and imaging capability of this survey provides 
an opportunity of discovery and localisation of pulsars and transients. Utilising this we localised 4 of the newly discovered 
pulsars in the gated image plane. With the aid of $\sim$ 10\arcsec~localisation, we obtained the timing solutions of 2 of 
the discoveries with relatively short data spans ($<$ 1/2 year). 
In addition to the regular emission from pulsars, the GHRSS survey will reveal transient events like the RRATs or the FRBs. With 
an ongoing enhancement of processing power we expect more results soon.

\acknowledgments
B. Bhattacharyya acknowledges support of Marie Curie grant PIIF-GA-2013-626533 of European Union. 
P. Ray's contributions to this work were supported by the Chief of Naval Research (CNR).
We thank the reviewer of our paper for a detailed review that improved the paper significantly.
We thank A. Holloway and R. Dickson for making the {\it Hydrus} computing cluster at University of Manchester available for this survey analysis. We also thank them in helping 
to set-up the GPU cluster. We also acknowledge the generous support of V. Venkatasubramani, S. J. Bhachal and A. Meghe at the NCRA in 
keeping the {\it IBM} cluster running for data processing. We thank L. Levin for {\sc pspoppy} related discussion and M. Mickaliger for plotfil display tool. 
We also thank E. Keane for a discussion on the paper.  
We acknowledge the help from N. Mohan for modifying the {\sc pybdsm} package to suit the imaging analysis of the GHRSS fields.
We acknowledge support of GMRT operators. The GMRT is run by the National Centre for Radio Astrophysics of the Tata Institute 
of Fundamental Research. The National Radio Astronomy Observatory is a facility of the National Science Foundation operated under cooperative
agreement by Associated Universities, Inc.


\end{document}